\documentstyle[prb,aps,amssymb,twocolumn,psfig,latexsym]{revtex}

\begin{document}

\draft

\twocolumn[\hsize\textwidth\columnwidth\hsize\csname
@twocolumnfalse\endcsname
\title{Thermodynamics and Spin Tunneling Dynamics 
in Ferric Wheels with Excess Spin}
\author{Florian Meier and Daniel Loss}
\address{Department of Physics and Astronomy, University of Basel, 
Klingelbergstrasse 82, 4056 Basel, Switzerland}
\date{July 2, 2001}

\maketitle

\begin{abstract}
We study theoretically the thermodynamic properties and spin dynamics of 
a class of magnetic rings closely related to ferric 
wheels, antiferromagnetic ring systems, in which one of the Fe (III) ions
has been replaced by a dopant ion to create an excess spin. 
Using a coherent-state spin path integral formalism, we derive an effective 
action for the system in the presence of a magnetic field. We calculate the
functional dependence of the magnetization and tunnel splitting on the
magnetic field and show that the parameters of the spin Hamiltonian
can be inferred from the magnetization curve. We study the spin dynamics
in these systems and show that
quantum tunneling of the N{\'e}el  vector also results in tunneling of the 
total magnetization. Hence, the spin correlation function 
shows a signature of N{\'e}el  vector tunneling, and 
electron spin resonance (ESR) techniques or AC susceptibility measurements
can be used to measure both the tunneling and the decoherence rate. 
We compare our results with exact diagonalization studies on small ring 
systems. Our results can be easily generalized to a wide class of nanomagnets,
such as ferritin.
\end{abstract}

\vskip2pc]

\section{Introduction}
\label{sec:introduction}

Nanomagnets and molecular clusters are 
systems in which macroscopic quantum phenomena may be observed 
in the form of quantum
tunneling of the magnetization.~\cite{korenblit:78,chudnovsky:88,magtunnel}
Two scenarios must be carefully distinguished: incoherent macroscopic quantum
tunneling (MQT) and macroscopic quantum coherence (MQC). 
In the latter case, tunneling between energetically degenerate spin
configurations takes place at a rate $\Delta/h$ large compared to
the spin decoherence rate $\Gamma$.
In ferromagnetic molecular clusters such as Fe$_8$ and 
Mn$_{12}$ the ground state tunnel splitting $\Delta$ is small compared
to $\hbar \Gamma$.~\cite{friedman:96,thomas:96,wernsdorfer:99} However, 
$\Delta$ is  
significantly larger in antiferromagnetic (AF) 
systems~\cite{barbara:90,krive:90} which are 
promising candidates for the observation of MQC in the form of
coherent tunneling of the N{\'e}el vector ${\bf n}$. 

\begin{figure}[f]
\centerline{\psfig{file=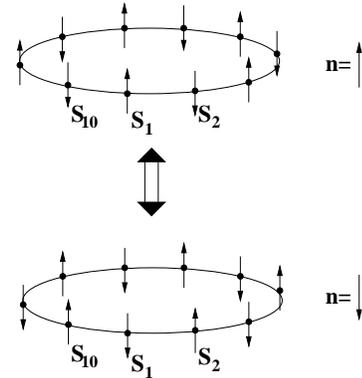,height=5.0cm}}
\caption{The two degenerate classical ground-state spin configurations 
of the ferric wheel Fe$_{10}$.}
\label{fig1}
\end{figure}

The ferric wheels (FWs) Li:Fe$_6$, Na:Fe$_6$, Cs:Fe$_8$, and 
Fe$_{10}$~\cite{gatteschi:94,taft:94,caneschi:96,waldmann:99,waldmann:00,chiolero:98} are 
a  particularly interesting class of molecular magnets. The 
$s=5/2$ Fe (III) ions are arranged on a ring, with an AF nearest-neighbor 
exchange coupling $J>0$, and a weak, easy-axis anisotropy ($k_z$) 
directed along the ring axis ${\bf e}_z$.
For 
$h_x=0$, the classical ground-state spin configuration has
alternating (N{\'e}el) order with the spins pointing 
along $\pm {\bf e}_z$. The two states 
with the N{\'e}el vector ${\bf n}$ along $\pm{\bf e}_z$ [Fig.~\ref{fig1}], 
labeled $|\uparrow \rangle$ and $|\downarrow \rangle$, are 
energetically degenerate and separated by an energy 
barrier of height  $N k_z s^2$. However, $|\uparrow \rangle$ and 
$|\downarrow \rangle$ are not energy eigenstates. Rather,
the low-energy sector of the FW 
consists of a ground state  $|g\rangle =(|\! \uparrow\rangle + 
| \! \downarrow\rangle )/\sqrt{2}$ 
and a first excited state $|e\rangle =(|\! \uparrow\rangle - | \! \downarrow
\rangle )/\sqrt{2}$, separated in energy by $\Delta$. For weak tunneling, 
$|g\rangle$ and $|e\rangle$
are energetically well separated from all other energy eigenstates. 

For Fe$_{10}$, the tunnel splitting $\Delta$ can be as large as 
$2.18$K.~\cite{chiolero:98,normand:00} 
An estimate for the electron spin
decoherence rate $\Gamma$ in FWs can be obtained from the typical energy scales
of the various interactions of the electron spins. These include nuclear
dipolar interactions with $^1$H nuclei 
($0.1$mK [Ref.~\onlinecite{cornia:00}]), 
hyperfine interactions with $^{57}$Fe ($1$mK [Ref.~\onlinecite{evans}]), and
interring electron spin dipolar interactions (of order $10$-$50$mK),
and possible interring superexchange processes, which are 
difficult to estimate but may be of order $100$mK. However, the tunnel 
splitting is sufficiently large that FWs are most promising candidates 
for coherent tunneling of ${\bf n}$ even in the presence of intrinsic or 
extrinsic sources of decoherence on energy scales up to $0.5$K.

Although quantum tunneling effects in antiferromagnets are 
more pronounced than in ferromagnets,~\cite{barbara:90,krive:90} the 
detection of 
quantum behavior is experimentally more challenging. The reason for 
this is that magnetization and susceptibility measurements probe only 
the total spin of the molecule which, by symmetry, remains unaltered 
upon tunneling of the N\'eel vector.
This problem is resolved in ferrimagnets and antiferromagnets with 
uncompensated sublattice spins, in which $\Delta$ is still large and
tunneling of ${\bf n}$ leads to large signals in the alternating-current (AC) 
susceptibility,~\cite{awschalom:92,awschalom:92b,gider:95} provided that 
magnetic fields are 
small.~\cite{loss:92,duan:95,chudnovsky:95,chiolero:97} Recent 
work~\cite{ivanov:99} indicates that ferrimagnets exhibit a wealth of 
interesting tunnel scenarios also in finite magnetic fields.

In this paper we study magnetic rings closely related to the ferric
wheels, in which one of the Fe (III) ions
has been replaced by a Ga or Cr ion with spin $s^\prime=0$ or
$s^\prime=3/2$, respectively,  to create an excess spin. Such systems have
been synthesized recently.~\cite{awschalom:private} Starting from 
a microscopic model Hamiltonian, 
\begin{eqnarray}
&& \hat{H} = J \sum_{i=2}^{N-1} \hat{{\bf s}}_i \cdot \hat{{\bf s}}_{i+1} + 
J^\prime ( \hat{{\bf s}}_1 \cdot \hat{{\bf s}}_{2} + \hat{{\bf s}}_1 
\cdot \hat{{\bf s}}_{N} )
\nonumber \\
&& + \hbar {\bf h} \cdot  
\sum_{i=1}^N \hat{{\bf s}}_i - (k_z^\prime \hat{s}_{1,z}^2 +
k_z   \sum_{i=2}^{N} \hat{s}_{i,z}^2),
\label{eq:1-h}
\end{eqnarray}
which also accounts for modified exchange ($J^\prime$) and 
anisotropy ($k_z^\prime$) constants due to doping, we calculate
various thermodynamic quantities and spin correlation functions. 
In Eq.~(\ref{eq:1-h}), $N=10$, $8$, or $6$, ${\bf h} = 
g \mu_B {\bf B}/\hbar$, with ${\bf B}$ the external magnetic field, and $g=2$ 
is the electron spin $g$-factor. 
As we will show, the excess spin $\delta s = s^\prime - s$ 
is strongly
coupled to ${\bf n}$ and hence  is
expected to modify both the
thermodynamic properties and the spin dynamics of the FW.  
In contrast,  for an impurity
spin coupled weakly to the N{\'e}el vector,~\cite{meier:01} the 
thermodynamic properties of the wheel remain essentially unaltered. 
For the modified FWs discussed in the present work, 
N{\'e}el vector tunneling also leads to 
oscillations of the total spin which are in principle observable in AC
susceptibility or ESR measurements. Thermodynamic
properties of AF systems with uncompensated sublattice spins by now have been 
studied in great detail for various anisotropy potentials and field 
configurations.~\cite{lu:97,lu:00} One main advantage of the small, 
high-symmetry modified FW  studied in the present work
is that the dependence of various thermodynamic quantities and spin
correlation functions on the small number of microscopic parameters entering
Eq.~(\ref{eq:1-h}) can be evaluated analytically.

The outline of this paper is as follows. In Sec.~\ref{sec:0-fw} we review the
theory of spin tunneling in FWs  
and calculate the spin correlation functions 
$\chi_{\alpha \alpha}$. In Sec.~\ref{sec:subl} we
discuss the modified FW within a classical vector model and show that, in 
contrast to the FW, tunneling of ${\bf n}$ now also results in  oscillations
of the total spin. In Sec.~\ref{sec:thermodyn} and Sec.~\ref{sec:dyn}, we 
develop a semiclassical theory of the modified FW, thus generalizing earlier 
work on AF systems with uncompensated sublattice 
spins~\cite{loss:92,duan:95,chudnovsky:95,chiolero:97} to 
$B_x \neq 0$.
For $|J^\prime/J-1| \ll 1$, we evaluate explicitly the tunnel splitting 
$\tilde{\Delta}$ and magnetization $M_x $ of the modified FW, and show that 
$J^\prime$ can be determined by measuring $M_x $. We calculate the Fourier
transform of the real-time susceptibility $\chi^{\prime
\prime}_{zz} (\omega \simeq \tilde{\Delta}/\hbar)$ and prove that N{\'e}el
vector tunneling can be detected in AC susceptibility or ESR measurements.
In Sec.~\ref{sec:TD-SD2}, we discuss the modified FW for the
limiting cases $J^\prime/J \gg 1$ and $J^\prime/J\ll 1$. The relevance of
the present work for experiments is discussed in Sec.~\ref{sec:discussion}.
Finally, we indicate that our results can be easily generalized to other 
systems
such as ferritin (Sec.~\ref{sec:ferritin}). We summarize
our results in Sec.~\ref{sec:conclusions}.

\section{Thermodynamics and Spin Dynamics of the FW}
\label{sec:0-fw}

\subsection{Thermodynamics}
\label{sec:0-mqc}

In this section, we give a brief review of previous work on N{\'e}el vector 
tunneling in FWs.~\cite{chiolero:98}
In particular, we point
out that the notion of quantum tunneling applies only to Fe$_{10}$ and
Cs:Fe$_8$ (and
somewhat less to Na:Fe$_6$) for large $h_x$, but not to Li:Fe$_6$. 

The minimal Hamiltonian of the FWs contains only two parameters, 
$J$ and $k_z$,
\begin{equation}
\hat{H}_0 = J \sum_{i=1}^N \hat{{\bf s}}_i \cdot \hat{{\bf s}}_{i+1} + 
\hbar {\bf h} \cdot  
\sum_{i=1}^N \hat{{\bf s}}_i - k_z   \sum_{i=1}^N \hat{s}_{i,z}^2.
\label{eq:1-h0}
\end{equation}
Here $N=10$, $8$, or $6$ and $\hat{{\bf s}}_{N+1} \equiv \hat{{\bf s}}_1$, 
${\bf h} = 
g \mu_B {\bf B}/\hbar$, with ${\bf B}$ the external magnetic field, and $g=2$ 
is the electron spin $g$-factor. 
Throughout this paper we restrict our
attention to ${\bf B}=B_x {\bf e}_x$, i.e. magnetic fields applied in the
ring plane.
For $k_z = 0$, the eigenstates of the total spin $\hat{{\bf S}}=\sum_{i=1}^N 
\hat{{\bf s}}_i$ are also energy eigenstates, with 
energy~\cite{taft:94,chiolero:98} 
$E_{S,S_x}\simeq(2J/N)S(S+1)+ \hbar h_x S_x$. 
For systems with weak anisotropy,
$k_z \ll 2 J/(Ns)^2$, the anisotropy can be taken into account
in perturbation theory for a wide range of $h_x$. However, the scenario 
changes for large anisotropy  
$k_z \gtrsim 2 J/(Ns)^2$, where mixing of 
different spin multiplets becomes appreciable, as is the case for 
Fe$_{10}$, Cs:Fe$_8$, and Na:Fe$_6$. 

Both the 
tunnel splitting $\Delta$ and the magnetization $M_x$ of the FW can 
be obtained from the partition function $Z_0[h_x]$ which we evaluate 
using spin path integrals. Introducing spin coherent states, we  
decompose the local spin fields ${\bf s}_i$~\cite{remark1c}
\begin{equation}
{\bf s}_i = (-1)^{i+1} s{\bf n} + {\bf l}
\label{eq:1-staggering}
\end{equation} 
into a N{\'e}el ordered field $\pm s{\bf n}$ (${\bf n}^2 =1$) and 
fluctuations ${\bf l}\perp {\bf n}$ around it. For small systems containing 
$6$, $8$, or
$10$ spins, spatial fluctuations of the fields ${\bf n}$ and ${\bf l}$ 
are frozen out at low temperature T. Carrying out the Gaussian integral over 
${\bf l}$, we obtain 
$Z_0 =\int {\mathcal D}{\bf n} \, \exp
\left[-\int_0^{\beta \hbar} 
{\rm d}\tau \, L_0[{\bf n}]/\hbar \right]$ with a Euclidean Lagrangean 
depending only on ${\bf n}$,
\begin{equation}
L_0[{\bf n}] = \frac{N \hbar^2}{8J} [-(i {\bf n} \times \dot{\bf n} 
- {\bf h})^2 + 
({\bf h}\cdot {\bf n})^2 - \omega_0^2 n_z^2],
\label{eq:1-action}
\end{equation} 
where $\omega_0 = s \sqrt{8 J k_z}/\hbar$.

In contrast to the classical description of the states 
$|\uparrow\rangle$ and $|\downarrow\rangle$ used in 
Sec.~\ref{sec:introduction}, in a quantum mechanical treatment ${\bf n}$ 
always exhibits quantum  fluctuations around its classical minima.
The notion of quantum tunneling, however, is only applicable if ${\bf n}$ is
well localized in the  states $|\!\uparrow\rangle$ and $|\!\downarrow\rangle$,
i.e. if  $1-\langle \uparrow\!| n_z^2 |\!\uparrow \rangle \ll 1$.
The  quantum  fluctuations of ${\bf n}$ can be 
estimated  from Eq.~(\ref{eq:1-action}). For ${\bf h}=0$, $L_0$ describes 
two independent harmonic oscillators of frequency $\omega_0$, corresponding to 
fluctuations of ${\bf n}$ in the direction of 
${\bf e}_x$ and ${\bf e}_y$. If the amplitude of the fluctuations is small,
we can evaluate the mean deviation of the N{\'e}el vector from ${\bf e}_z$, 
$1-\langle n_z^2 \rangle \simeq 2/({\mathcal S}_0/\hbar)$, where
${\mathcal S}_0/\hbar = N s \sqrt{2 k_z/J}$ is the classical tunnel action. 
Hence, ${\bf n}$ 
is well localized along ${\bf e}_z$ only if ${\mathcal S}_0/\hbar \gg 2$ or, 
equivalently, if the ground-state energy, 
$2 \times \omega_0/2$, is small compared to the potential barrier $N k_z s^2$.
The scenario changes if a strong magnetic field ${\bf h}= h_x 
{\bf e}_x$, $h_x \gg \omega_0$,  is applied in the ring plane. Then, the
mode of ${\bf n}$ along ${\bf e}_x$ is frozen out, such that 
$1-\langle n_z^2 \rangle \simeq 1/({\mathcal S}_0/\hbar)$, and the FW can exhibit
quantum tunneling if ${\mathcal S}_0/\hbar \gg 1$. Note that for large
tunnel action, ${\mathcal S}_0/\hbar \gtrsim 10$, the tunnel splitting becomes
small, which would make the system under consideration a less favorable
candidate for the observation of MQC.

The FWs Li:Fe$_{6}$, Na:Fe$_{6}$, Cs:Fe$_8$, and Fe$_{10}$ have been well 
characterized.~\cite{taft:94,caneschi:96,waldmann:99,waldmann:00,normand:00,cornia:99} 
For Fe$_{10}$, $J=15.56 {\rm K}$ and 
$k_z = 0.0088 J$. 
For Cs:Fe$_8$, $J=22.5$K and $k_z=0.0191J$.~\cite{waldmann:00}
For Fe$_6$, $J$ and $k_z$ vary 
appreciably depending on the central alkali metal ion and ligands: 
for Na:Fe$_6$, 
$J=32.77 {\rm K}$ and $k_z = 0.0136 J$, whereas for  Li:Fe$_6$, $J = 
20.83 {\rm K}$ and $k_z = 0.0053 J$.~\cite{cornia:99,normand:00}
In Table~\ref{tab1}, ${\mathcal S}_0/\hbar$ and $\hbar \omega_0/2 \mu_B$ are 
given for Fe$_{10}$, Cs:Fe$_8$, Na:Fe$_{6}$, and Li:Fe$_6$. 
As is obvious from these
values, for none of the molecular rings ${\mathcal S}_0/\hbar$ is
sufficiently large to assure that a tunnel scenario is rigorously applicable 
if $h_x \lesssim \omega_0$. 
In Na:Fe$_6$, even at large $B_x \gg 20$T, ${\bf n}$ is far
less well localized along ${\bf e}_z$ than in Fe$_{10}$, which hence
remains the most favorable candidate for the observation of quantum
tunneling. Note however that even in Fe$_{10}$ and Cs:Fe$_8$, 
${\mathcal S}_0/\hbar$ is so
small that corrections to the instanton techniques used below may become
large.~\cite{remark1} 

\begin{table}[!f]
\begin{tabular}{|l|c|c|}
& ${\mathcal S}_0/\hbar$ & $\hbar \omega_0/2 \mu_B$ \\
\hline
Fe$_{10}$ & $3.32$ & $ 7.68$ T \\
Cs:Fe$_8$ & $3.91$ & $ 16.37$T \\
Na:Fe$_6$ & $2.47$ & $ 20.11$T \\
Li:Fe$_6$ & $1.54$ & $ 7.98$T 
\end{tabular}
\caption{${\mathcal S}_0/\hbar=Ns \sqrt{2 k_z/J}$ and $\hbar \omega_0/2 \mu_B$ 
for  Fe$_{10}$, Cs:Fe$_8$, Na:Fe$_{6}$, and Li:Fe$_6$. }
\label{tab1}
\end{table}

For $h_x \gg \omega_0$, the magnetic field 
${\bf B}$ strongly confines ${\bf n}$ to the $(y,z)$-plane and thus determines
the tunnel path of the electron spins. This allows one to 
evaluate $Z_0$. We parameterize
\begin{equation}
{\bf n}=(\cos \theta, \sin \theta \, \cos \phi, \sin \theta \, \sin \phi)
\label{eq:2-param}
\end{equation}
and expand $L_0$ to second order in $\vartheta=\theta-\pi/2$, 
\begin{equation}
L_0[{\bf n}] = \frac{N \hbar^2}{8J}[-( h_x -i\dot{\phi})^2 
- \omega_0^2 \sin^2 \phi] + 
\frac{1}{2} \vartheta G^{-1}[\phi] \vartheta,
\label{eq:2-action}
\end{equation}
where $G^{-1}[\phi] = (N \hbar^2/4J)(-\partial_\tau^2 + ( h_x -i\dot{\phi})^2 
 + \omega_0^2 \sin^2 \phi)$, $\dot{\phi}=\partial_\tau \phi$, and 
$\omega_0=s \sqrt{8 J k_z}/\hbar$ as defined above. The typical energy scales 
for the dynamics of
$\phi$ and $\vartheta$ are $\omega_0$ and $h_x$, respectively. Due to this 
separation of energy-scales, we can use an adiabatic approximation, in which 
$\vartheta$ oscillates rapidly
in a quasistatic harmonic potential~\cite{remark1a} 
$(N \hbar^2/8J)(h_x^2-2 i h_x \dot{\phi})$. 
Integrating 
out $\vartheta$, we obtain an expression for $L_0$ depending only on 
$\phi$.~\cite{chiolero:98} For $k_B T/\hbar \ll \omega_0 \ll h_x$, we find
\begin{eqnarray}
L_0 [\phi] & \simeq &\frac{N \hbar^2}{8J} [-( h_x -i\dot{\phi})^2 
- \omega_0^2 \sin^2 \phi] + \hbar \frac{h_x-i \dot{\phi}}{2} \nonumber \\
&& \hspace*{2cm}+ {\mathcal O}
(\omega_0^2/h_x),
\label{eq:2-actionphi}
\end{eqnarray}
where the term $\hbar(h_x-i \dot{\phi})/2$ arises from the 
$\vartheta$ fluctuation
determinant. The two saddle-points of Eq.~(\ref{eq:2-actionphi}),
$\phi \equiv \pi/2$ and $\phi \equiv 3 \pi/2$, correspond to the
two classical spin configurations sketched in Fig.~\ref{fig1}. If tunneling
is weak, ${\mathcal S}_0/\hbar \gg 1$,
the remaining path integral over $\phi$ is 
straightforward.~\cite{kleinert,chiolero:98} 
Summing all multi-instanton solutions, one finds 
\begin{equation}
Z_0=\exp \left[\beta \left(\frac{N \hbar^2}{8J} h_x^2 - \hbar \frac{h_x + 
\omega_0}{2} \right)\right]
\cosh \left(\frac{\beta \Delta(h_x)}{2}\right)
\label{eq:2-partitionf}
\end{equation}
with the tunnel splitting
\begin{equation}
\Delta  (h_x) = \Delta_0 \left|\sin \left( 
\pi \frac{ N \hbar}{4J} h_x  \right) \right|,
\label{eq:delta}
\end{equation}
where $\beta = 1/k_B T$, $\Delta_0 = 8 \hbar \omega_0 \sqrt{{\mathcal S}_0/2 
\pi\hbar} \, \exp[-{\mathcal S}_0/\hbar]$, and ${\mathcal S}_0/\hbar=
Ns\sqrt{2k_z/J}$. In particular, 
$\Delta$ is periodic as a function of $h_x$. Differentiating
with respect to $B_x$, we obtain the magnetization~\cite{chiolero:98}
\begin{equation}
M_x = (g \mu_B) \left[ \frac{N \hbar}{4J} h_x -\frac{1}{2} 
+ \frac{1}{2 \hbar} \frac{\partial \Delta}{\partial h_x} \tanh \left(
\frac{\beta \Delta}{2} \right)
\right]  . 
\label{eq:2-magx}
\end{equation}
Because $\partial \Delta/\partial h_x$ is discontinuous at the zeroes of 
$\Delta$, $M_x$ exhibits steps at $B_{c,n} = n 4 J/N g \mu_B$, where 
$n=1,2,\ldots, Ns$. From
\begin{eqnarray}
M_\alpha & = & (g \mu_B) \frac{N \hbar}{4J} \frac{1}{Z}
\int {\mathcal D}{\bf n} \, 
[ h_\alpha - 
i ({\bf n}\times \dot{\bf n})_\alpha - n_\alpha {\bf h}\cdot {\bf n} ]
\nonumber  \\ && \hspace*{2.5cm}
\times e^{-\int_0^{\beta \hbar}{\rm d}\tau \, L_0 [{\bf n}]/\hbar},
\label{eq:2-magy}
\end{eqnarray}
it also follows that, for arbitrary $h_x$, $M_\alpha=0$ for 
$h_\alpha=0$, $\alpha=y,z$, which is a result of the invariance
of $\hat{H}_0$ under rotation around ${\bf B}$ by $\pi$.~\cite{remark1b}

\subsection{Spin dynamics}
\label{sec:0-dyn}

We show now that the tunneling dynamics of ${\bf n}$ does not enter
the susceptibility of the FW, i.e. that the 
imaginary part of the susceptibility, 
$\chi^{\prime \prime}_{\alpha \alpha}(\omega)$, has no 
absorption peak $\delta (\omega-\Delta/\hbar)$. To prove this,
 we calculate the susceptibility of the total spin, 
$\hat{{\bf S}}= 
\sum_{i=1}^N \hat{{\bf s}}_i$, in Matsubara representation, 
\begin{eqnarray}
\chi_{\alpha \alpha} (i \omega_n) &= & (g \mu_B)^2 \int_0^{\beta \hbar} 
{\rm d}\tau \, e^{i \omega_n \tau}
\bigl[ \langle T_\tau \hat{S}_\alpha (\tau) \hat{S}_\alpha (0) 
\rangle \nonumber \\  && \hspace*{3.5cm} - \langle  \hat{S}_\alpha  \rangle^2
\bigr],
\label{eq:2-imagtimesusc}
\end{eqnarray}
where $\alpha=x,y,z$. As we will show below, 
$\chi_{\alpha \alpha} (i \omega_n)$ contains {\it no} terms
proportional to $1/(i \omega_n - \Delta/\hbar)$. 
This implies that AC susceptibility or ESR 
measurements cannot be used to detect 
N{\'e}el vector tunneling in FWs described by $\hat{H}_0$.

$\chi_{xx} (i \omega_n)$ can be calculated from the  
generating functional $Z[\delta h_x(\tau)]$, where 
$\delta h_x(\tau)$ is a small probing field added to the static field $h_x 
\gg \omega_0$. 
Because we are only interested in the low frequency response of the FW, 
$\omega_n \lesssim \Delta/\hbar \ll \omega_0$, we may
restrict our attention to a slowly varying field $\delta h_x$ whose Fourier 
components vanish for $\omega_n \gtrsim \Delta/\hbar$. The typical 
timescale for dynamics of $\phi$, $1/\omega_0$, is short compared
to the timescale on which $\delta h_x$ varies. In particular, approximating
$\delta h_x (\tau)$ by a constant during instanton passage, we find 
\begin{eqnarray}
&& Z[\delta h_x (\tau)] \simeq \exp[\int_0^{\beta \hbar} 
\frac{{\rm d}\tau}{\hbar} \,\left( 
\frac{N \hbar^2}{8J} (h_x + \delta h_x(\tau) )^2 \right. \label{eq:2-partfb} 
\\ &&
\left. - \hbar \frac{h_x + \delta h_x(\tau) +\omega_0 }{2} \right)]
\cosh [\int_0^{\beta \hbar} \frac{{\rm d}\tau}{\hbar} \,
\frac{\Delta (h_x + \delta h_x (\tau))}{2}]. 
\nonumber 
\end{eqnarray}
Differentiating twice and setting $\delta h_x \rightarrow 0$, for
$\omega_n \lesssim \Delta/\hbar$,
\begin{eqnarray}
\chi_{xx}(i \omega_n) &=&  (g \mu_B)^2 
 \Bigl[\frac{N \hbar}{4J} + \frac{1}{2 \hbar}
\frac{\partial^2 \Delta}{\partial h_x^2} \tanh(\beta \Delta/2)
\nonumber \\
&&
+ \frac{\beta}{\hbar} \left(\frac{1}{2}\frac{\partial \Delta}{\partial h_x}
\right)^2
\cosh^{-2}(\beta \Delta/2) \delta_{\omega_n,0} \Bigr].
\label{eq:2-suscx}
\end{eqnarray}

The transverse susceptibilities can be evaluated directly from
$\delta^2 Z/\delta h_\alpha (\tau) \, \delta h_\alpha
(0)$, $\alpha =y,z$. Using the parametrization in 
Eq.~(\ref{eq:2-param}), 
$\vartheta = \theta-\pi/2$ can be integrated out in the path-integral
expression for $\chi_{\alpha 
\alpha}$. After lengthy calculation (Appendix~\ref{sec:a-suscept}), 
we obtain for $\omega_n \ll h_x$ 
\begin{eqnarray}
\chi_{yy}(i \omega_n) & \simeq & (g \mu_B)^2 \frac{N \hbar}{4J}
, \nonumber \\
\chi_{zz}(i \omega_n) & \simeq & (g \mu_B)^2 \frac{N \hbar}{4J}
. \label{eq:2-suscy}
\end{eqnarray}

From Eqs.~(\ref{eq:2-suscx}) and (\ref{eq:2-suscy}) it is evident that 
none of the susceptibilities $ \chi_{\alpha \alpha}
(i \omega_n)$ contains a term proportional to 
$1/(i \omega_n \pm \Delta/\hbar)$.
In the tunneling regime discussed here, $|g\rangle$ and $|e\rangle$ are
energetically well separated from all other states. Then, 
Eq.~(\ref{eq:2-suscy}) and the spectral representation 
\begin{eqnarray}
\chi_{\alpha \alpha}(i \omega_n) &=& \sum_{i,j} \frac{e^{-\beta E_i}}{Z}
|\langle i | \hat{S}_\alpha | j \rangle |^2 \left( \frac{1}{i \omega_n
- (E_i - E_j)/\hbar } \right. \nonumber \\ && \left.
- \frac{1}{i \omega_n + (E_i - E_j)/\hbar } 
\right) - \beta \hbar \delta_{\omega_n,0} \langle \hat{S}_\alpha \rangle^2
\label{eq:2-chispectral}
\end{eqnarray}
where $|i \rangle$ and $|j \rangle$ label energy eigenstates, 
imply that $\langle e| \hat{S}_\alpha | g \rangle = 0$ ($\alpha 
= x,y,z$).  
Although for the parameters of Fe$_{10}$, Cs:Fe$_8$, 
and Na:Fe$_6$ tunnel corrections to $\chi_{\alpha
\alpha}$ neglected in Eq.~(\ref{eq:2-suscy}) can be significant, the main 
conclusion of our calculation -- that coherent tunneling of ${\bf n}$ does 
not enter the susceptibilities $\chi_{\alpha \alpha}$ -- remains valid. 

Indeed, $\langle e | \hat{S}_\alpha |g \rangle =0$ is a direct consequence of 
the invariance of $\hat{H}_0$ as $i \rightarrow i+1$, i.e. the
exchange of the two sublattices of the bipartite AF ring. In order to clarify
this point, we introduce the sublattice spin operators
\begin{equation}
\hat{{\bf S}}_A = \sum_{i={\rm odd}} \hat{{\bf s}}_i, \hspace*{0.5cm}
\hat{{\bf S}}_B = \sum_{i={\rm even}} \hat{{\bf s}}_i. 
\label{eq:2-sublspins}
\end{equation} 
with $\hat{\bf S} = \hat{{\bf S}}_A + \hat{{\bf S}}_B$.
In a semiclassical description of the FW, spins of one sublattice couple
ferromagnetically to each other. The classical 
spin fields~\cite{remark1c} ${\bf s}_i$ then obey ${\bf s}_i =
{\bf S}_A/(N/2)$ for odd $i$, and  ${\bf s}_i = {\bf S}_B/(N/2)$  for even $i$.
This amounts to
treating ${\bf S}_{A}$ and ${\bf S}_B$ as single large 
spins~\cite{barbara:90,krive:90,duan:95} with spin 
quantum number $N s/2$, and $\hat{H}_0$ reduces to
\begin{eqnarray}
\hat{H}_{0,{\rm subl}} &=& \frac{4J}{N} \hat{{\bf S}}_A \cdot \hat{{\bf S}}_B 
+ \hbar {\bf h}\cdot 
(\hat{{\bf S}}_A +  \hat{{\bf S}}_B) \nonumber \\ 
&& \hspace*{1cm} - \frac{2 k_z}{N}  ( \hat{S}_{A,z}^2 + \hat{S}_{B,z}^2). 
\label{eq:2-sublham}
\end{eqnarray}
Similar to the 
nonlinear sigma model (NLSM)-formalism used above, $ \hat{H}_{0,{\rm subl}}$ 
provides an appropriate description of the low-energy physics of the FW.

\begin{figure}
\centerline{\psfig{file=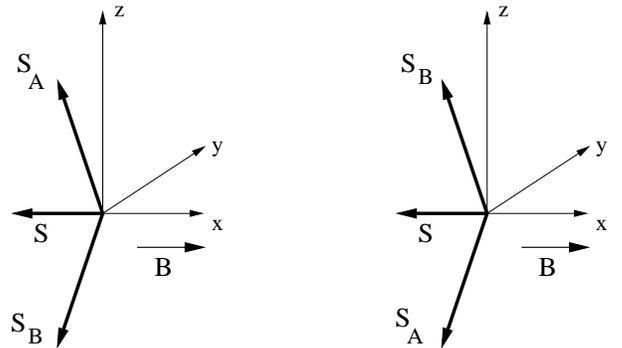,width=8.0cm}}
\caption{Sublattice spins and total spin of the FW in magnetic fields.}
\label{fig2}
\end{figure}

In Fig.~\ref{fig2}, the two classical ground-state spin configurations of 
$\hat{H}_{0,{\rm subl}}$ are shown. A finite magnetic field ${\bf B}$ tilts the
sublattice spins away from the direction of ${\bf B}$ such that 
${\bf S} = {\bf S}_A + {\bf S}_B$ is  parallel to ${\bf B}$.
During tunneling of ${\bf n}$, the sublattice spins retain their position 
relative to each other and rotate jointly around ${\bf e}_x$. 
The total spin vector ${\bf S}$, however, remains
invariant during tunneling such that the real-time spin correlation
function does not contain terms proportional to $e^{i \Delta t/\hbar}$ or, 
equivalently,
$\langle e | \hat{S}_\alpha | g \rangle = 0$.

\section{A phenomenological model for the modified FW}
\label{sec:subl}

The fact that the susceptibilities $\chi_{\alpha \alpha}$ of the FW show no
signature of the tunneling of ${\bf n}$ is a consequence of the symmetry 
of $\hat{H}_{0,{\rm subl}}$ under 
$\hat{{\bf S}}_A  \leftrightarrow \hat{{\bf S}}_B$. In the modified FWs 
[Eq.~(\ref{eq:1-h})] this symmetry is broken, which makes them 
much more suitable for the observation of tunneling of ${\bf n}$.
In this section we discuss the saddle-point, i.e. classical, properties of
the phenomenological sublattice model for the modified FW. 
We generalize earlier 
work~\cite{loss:92,duan:95,chudnovsky:95,chiolero:97} to finite magnetic 
fields and
show that, in contrast to systems with easy-plane anisotropy,~\cite{ivanov:99} 
in molecular rings with easy-axis anisotropy and finite excess spin, ${\bf S}$
oscillates as ${\bf n}$ tunnels. 

Following Sec.~\ref{sec:0-dyn}, we introduce $\hat{{\bf S}}_{A}$ and 
$\hat{{\bf S}}_{B}$ as 
defined in Eq.~(\ref{eq:2-sublspins}). The Hamiltonian 
$\hat{H}$ of the modified FW can thus be mapped onto a simpler version in 
terms of sublattice spins,
\begin{eqnarray}
&& \hat{H}_{\rm subl} = \frac{4[(N-2)Js^2 +2J^\prime s s^\prime]}{
(N s + 2 \delta s)N s} 
\hat{{\bf S}}_A \cdot \hat{{\bf S}}_B + \hbar {\bf h}\cdot 
(\hat{{\bf S}}_A +  \hat{{\bf S}}_B) \nonumber \\ && - \frac{2 k_z}{N} 
\left[ \frac{1+2(k_z^\prime s^{\prime 2}-k_z s^2)/(Nk_z s^2)}{
(1+2 \delta s/N s)^2} \hat{S}_{A,z}^2 + \hat{S}_{B,z}^2 \right], 
\label{eq:3-sublh}
\end{eqnarray}
where $\delta s= s^\prime - s$,  
$S_A = (N/2-1)s + s^\prime = Ns/2 + \delta s$, and $S_B = N s/2$. 
Throughout this paper we assume the following inequalities:
\begin{eqnarray}
|\delta s| & \ll & N s, \label{eq:3-assumpt1} \\
|k_z^\prime s^{\prime 2}-k_z s^2| & \ll & N k_z s^2/2, J s  
\label{eq:3-assumpt2}, \\
2 J^\prime s^\prime & \ll & N J s \label{eq:3-assumpt3},
\end{eqnarray}
where $J, J^\prime>0$.
Eq.~(\ref{eq:3-assumpt1}) guarantees that the modified FW is an AF system 
with small
excess spin, to which the theory of N{\'e}el vector tunneling applies.
Both for Cr and Ga dopant ions ($|\delta s|=1$ and $5/2$, 
respectively), Eq.~(\ref{eq:3-assumpt1}) is well
satisfied. Eq.~(\ref{eq:3-assumpt2}) will allow us to 
treat the difference in sublattice anisotropies, 
$(k_z^\prime s^{\prime 2}-k_z s^2)$, in perturbation theory. 
Typical values of $k_z^\prime$ and $k_z$ are on the order of only $0.01 J$, 
such that this condition holds for most systems of
interest. Finally, Eq.~(\ref{eq:3-assumpt3}) together with 
Eq.~(\ref{eq:3-assumpt1}) assures that 
the `bulk' parameters of a FW are only slightly altered
by exchanging one single spin, such that the parameters of the 
undoped FW [Table~\ref{tab1}] still determine whether the modified FW is in a 
quantum tunneling regime. However, as will be shown below, the excess 
spin $\delta s \neq 0$ leads to qualitative changes in both thermodynamic 
and dynamic quantities. We further assume
\begin{equation}
k_B T \ll \hbar \omega_0 ,
\label{eq:3-assumpt5}
\end{equation}
which allows us to restrict our attention to the low-energy sector of the FW,
which consists of two tunnel-split states only.

\begin{figure}
\centerline{\psfig{file=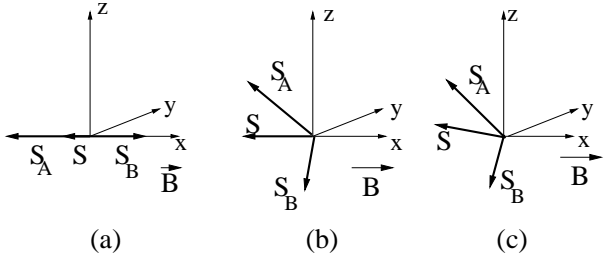,width=8.0cm}}
\caption{Classical ground-state spin configurations 
(sche\-ma\-tical\-ly) 
of an AF system with excess spin in a magnetic field. (a) $k_z=k_z^\prime=0$ 
and $h_x < h_c=|\delta s| 8 J/N \hbar$. 
(b) $h_x > h_c$ and $k_z=k_z^\prime=0$. (c) $h_x > h_c$. Now, the finite
anisotropies $k_z, k_z^\prime \neq 0$ lead to a finite component $S_z$
perpendicular to the magnetic field ${\bf B}$.}
\label{fig3}
\end{figure}

We first discuss the classical vector model of Eq.~(\ref{eq:3-sublh}) for 
$k_z = k_z^\prime = 0$, but finite $h_x$. For an AF system with equal 
sublattice spins, the spins would lie close to the 
plane perpendicular to the field $h_x$. As sketched
in Fig.~\ref{fig2}, tilting of the spins  leads to a
gain in energy
$N \hbar^2 h_x^2/8 J$. However, for uncompensated sublattices, the 
configuration sketched in Fig.~\ref{fig3}a provides an energy gain   
$ |\delta s| \hbar
 \, h_x$ and hence is favorable for $\hbar h_x \ll |\delta s|\, 8 J/N= \hbar 
h_c$. Only for $h_x \gg h_c$, the classical ground-state 
spin configuration is as sketched in Fig.~\ref{fig3}b. The energy is minimized
if the projections of ${\bf S}_A$ and ${\bf S}_B$ onto the $(y,z)$-plane are 
antiparallel, such that ${\bf S}={\bf S}_A + {\bf S}_B$ is parallel to 
${\bf B}$. This picture remains valid for a system with 
easy-{\it plane} anisotropy and ${\bf B}$ applied
in the easy plane.~\cite{ivanov:99} 

The scenario changes for easy-{\it axis} (${\bf e}_z$) anisotropy and 
a magnetic field ${\bf B}$ perpendicular to ${\bf e}_z$. 
First, the anisotropy favors the spin configuration sketched in
Fig.~\ref{fig3}b over that  in Fig.~\ref{fig3}a. The true classical 
ground-state spin configuration depends on the ratio $k_z/\hbar h_x$. 
More important, even for $h_x \gg h_c$, ${\bf S}$ now has a component 
perpendicuar to ${\bf B}$. The reason for this is that for $\delta s \neq 0$ or
$k_z^\prime s^{\prime 2}-k_z s^2 \neq 0$, Eq.~(\ref{eq:3-sublh}) is no longer
invariant under exchange of ${\bf S}_A$ and ${\bf S}_B$ if $k_z \neq 0$. 
Due to Eq.~(\ref{eq:3-assumpt2}) and (\ref{eq:3-assumpt3}), 
the components $S_{y}$ and $S_z$ of the total spin 
can be evaluated perturbatively. With the polar angle
$\phi$ parameterizing the projection of ${\bf S}_A$ onto the $(y,z)$-plane, 
to leading order in $\delta s$ and 
$(k_z^\prime s^{\prime 2} - k_z s^2)/2 J s$ we obtain 
\begin{equation}
\left( \begin{array}{c} S_y \\ S_z \end{array} \right) \simeq 
\left( \delta s \, 
\frac{\omega_0^2}{h_x^2} + \frac{k_z^\prime s^{\prime 2}
- k_z s^2}{2 J s} \right) \, \sin^2 \phi \, 
\left( \begin{array}{c} \cos \phi \\ \sin \phi \end{array} \right).
\label{eq:3-totspin}
\end{equation}
As is evident from Eq.~(\ref{eq:3-totspin}), finite $S_{y}$ or $S_{z}$ can 
be due 
to $\delta s \neq 0$ or $k_z^\prime s^{\prime 2}-k_z s^2 \neq 0$, i.e. 
excess spin or unequal effective anisotropies for ${\bf S}_A$ and ${\bf S}_B$.

According to Eq.~(\ref{eq:3-assumpt1}), the modified FW is an AF system with 
small excess spin which
is expected to exhibit spin tunnel dynamics qualitatively similar to
the FW, as indicated by the close formal analogy between 
Eqs.~(\ref{eq:2-sublham}) and (\ref{eq:3-sublh}). For magnetic fields 
\begin{equation}
{\rm max}[\hbar \omega_0, |\delta s| 8J/N] \ll \hbar h_x \ll 4 Js,
\label{eq:3-assumpt4}
\end{equation}
the sublattice spin vectors ${\bf S}_{A}$ and ${\bf S}_{B}$
lie close to the $(y,z)$-plane. Due to the easy-axis anisotropy, configurations
with ${\bf S}_{A}$ and ${\bf S}_{B}$ close to the $z$-axis are energetically
favorable.
It is noteworthy that the condition $h_x \gg  \omega_0$ is not 
indispensible for quantum tunneling of ${\bf n}$, but only assures that a 
tunnel scenario remains applicable for a wider range of $k_z$ 
[Sec.~\ref{sec:0-mqc}]. In contrast, $\hbar h_x \gg |\delta s| 8J/N$ guarantees
that there are two energetically degenerate, macroscopically distinct 
spin configurations between
which spin tunneling may take place~\cite{ivanov:99} and hence, in general,
will shift the range of magnetic fields in 
which a tunneling scenario as discussed in the present context 
[Fig.~\ref{fig1}] is valid. 
Henceforth, we will always assume that $B_x$ is large enough to satisfy
Eq.~(\ref{eq:3-assumpt4}). For sufficiently
large $k_z$, the two-state model for the low
energy sector of the system outlined in Sec.~\ref{sec:introduction} then still
applies. As Eq.~(\ref{eq:3-totspin}) indicates, the modified FW 
exhibits one important novel feature: as ${\bf n}$ tunnels between 
${\bf e}_z$ and $-{\bf e}_z$ ($\phi = \pi/2$ and $\phi=3 \pi/2$,
respectively), the $z$-component of the total spin, $S_z$, oscillates 
between $S_0= (\delta s \, \omega_0^2/h_x^2 + (k_z^\prime s^{\prime 2}
- k_z s^2)/2 J s)$ and $-S_0$.

\section{Thermodynamics of the modified FW}
\label{sec:thermodyn}

In order to quantify the statements on N{\'e}el vector tunneling in modified
FWs [Sec.~\ref{sec:subl}], we now develop a semiclassical theory of
the modified FW. 
In this section, we discuss thermodynamic quantities such as the
magnetization $M_x$ and specific heat $c_v$. While our theory treats the 
spins semiclassically, we give the explicit dependence on the microscopic 
parameters of $\hat{H}$ [Eq.~(\ref{eq:1-h})]. To this end
we evaluate the partition function $Z$, 
thus generalizing the procedure reviewed in Sec.~\ref{sec:0-mqc} to systems 
with $k^\prime_z \neq k_z$, $J^\prime \neq J$, and $\delta s \neq 0$.  
The 
most significant change is that, for $\delta s \neq 0$, the staggering 
[Eq.~(\ref{eq:1-staggering})] must be
modified in order to account for ${\bf s}_1^2 = s^{\prime 2}$. The ansatz 
\begin{eqnarray}
{\bf s}_1 &=& s^\prime {\bf n} + \frac{s^\prime}{s} {\bf l}, \nonumber \\
{\bf s}_i & = & (-1)^{i+1} s {\bf n} + {\bf l} \hspace{1cm} \forall i \neq 1
\label{eq:3-staggering}
\end{eqnarray}
is equivalent to the assumption that spins within sublattices
$A$ and $B$ are ferromagnetically 
coupled.~\cite{barbara:90,krive:90,loss:92,duan:95,chudnovsky:95,chiolero:97}
The results for the 
magnetization $M_x$ and susceptibilities obtained from this ansatz
turn out to be in good agreement with those obtained from 
numerical exact diagonalization (ED) (see below and Sec.~\ref{sec:dyn}) as 
long as $|J^\prime/J -1| \ll 1$. We 
restrict ourselves to this case first and discuss further limiting cases
$J^\prime \ll J$ and $J^\prime \gg J$ in Sec.~\ref{sec:TD-SD2}.

As for the FW,~\cite{chiolero:98} at low temperatures spatial variations of 
the N{\'e}el  field ${\bf n}$, and the fluctuations ${\bf l}$ around it, are 
suppressed in small ring system. With the coherent spin states defined in 
Eq.~(\ref{eq:3-staggering}),~\cite{auerbach}  
\begin{equation}
Z = \int {\mathcal D}{\bf n} \,  {\mathcal D}{\bf l} \, \delta ({\bf n}\cdot
{\bf l}) e^{- S[{\bf n},{\bf l}]/\hbar},
\label{eq:3-za}
\end{equation}
where
\begin{eqnarray}
S[{\bf n},{\bf l}] &=& - i \hbar \left( s^\prime \omega[{\bf n}+ \frac{1}{s}
{\bf l}] + s \sum_{i=2}^{N} \omega[(-1)^{i+1}{\bf n}+ \frac{1}{s} {\bf l}] 
\right) \nonumber \\ && \hspace*{1cm} + 
\int_0^{\beta \hbar} {\rm d}\tau \,  H[{\bf n},{\bf l}].
\label{eq:3-act}
\end{eqnarray}
Here,
\begin{eqnarray}
&& H[{\bf n},{\bf l}] = 
2 [(N-2)J+2 \frac{s^\prime}{s}J^\prime]{\bf l}^2 
+ \hbar (N+\frac{\delta s}{s}) {\bf h}\cdot {\bf l} \nonumber \\ 
&& \hspace*{0.5cm}
- [(N-1) k_z s^2 + k_z^{\prime} s^{\prime 2}] n_z^2 + \frac{2}{s}(k_z s^2 
- k_z^\prime s^{\prime 2}) n_z l_z \nonumber \\ 
&& \hspace*{5cm}
+ \delta s \, \hbar {\bf h}\cdot{\bf n} 
\label{eq:3-h}
\end{eqnarray}
is the classical energy of a given spin configuration. 
The small term $[(N-1)k_z + k_z^\prime]l_z^2$ has already been neglected. 
The first term in Eq.~(\ref{eq:3-act}) is the Berry-phase term, where 
$\omega[\Omega]$ denotes the area traced out by some vector $\Omega(\tau)$ 
on the unit sphere. Here, we 
use the south pole gauge (i.e. $\omega[\Omega(\tau) 
\equiv -{\bf e}_x]=0$). Expanding the Berry-phase term to leading order in
${\bf l}$ with the parametrization Eq.~(\ref{eq:2-param}) yields
\begin{equation}
- \int_0^{\beta \hbar} {\rm d}\tau \, \Bigl[ i \delta s \, \dot{\phi}
(1+\cos \theta) 
+ i (N+\frac{\delta s}{s}) ({\bf n}\times \dot{\bf n})\cdot {\bf l} \Bigr].
\label{eq:3-berryphase}
\end{equation}
The first term is due to the fact that, for $\delta s 
\neq 0$, the Berry-phase terms of the AF ordered components $s^\prime 
\omega[{\bf n}] + s \sum_{i=2}^N \omega[(-1)^{i+1}{\bf n}]= \int d {\rm \tau}
(N s \dot{\phi}
+ \delta s \, \dot{\phi} (1+\cos \theta))$ do not add to an
integer multiple of $2 \pi$.~\cite{braun:96,loss:98}

Carrying out the Gaussian integral over ${\bf l}$ we obtain 
$Z= \int {\mathcal D}{\bf n} \, 
\exp[-\int_0^{\beta \hbar} {\rm d} \tau L[{\bf n}]/\hbar]$, with a Euclidean 
Lagrangean
\begin{eqnarray}
 L[{\bf n}] &=& \frac{N \hbar^2}{8 \tilde{J}} [-({\bf h} + h_A n_z {\bf e}_z
- i {\bf n}\times \dot{\bf n})^2  \nonumber \\ && \hspace*{0.5cm}
+ (({\bf h} + h_A n_z {\bf e}_z) 
\cdot {\bf n})^2 - \tilde{\omega}^2_0 n_z^2] \nonumber  \\ && \hspace*{0.5cm} 
+  \delta s \, \hbar [{\bf h}\cdot {\bf n} - i \dot{\phi} (1+\cos \theta)],
\label{eq:3-lagrange}
\end{eqnarray}
where 
\begin{eqnarray}
\frac{\tilde{J}}{N} & = & \frac{(N-2)J+2 s^\prime J^\prime/s}{(N + 
\delta s/s)^2}, \nonumber \\
h_A &=& \frac{2 (k_z s^2 - k_z^\prime s^{\prime 2})}{(N+\delta s/s)s \hbar }, 
 \label{eq:3-newparameters} \\
\tilde{\omega}^2_0 & = & \frac{8 \tilde{J}}{N}[(N-1)k_z s^2 + k_z^\prime 
s^{\prime 2}]. \nonumber 
\end{eqnarray}
Eq.~(\ref{eq:3-lagrange}) is the analog of Eq.~(\ref{eq:1-action}) for the 
FW.
A finite excess spin $\delta s$ of the modified FW leads to two significant
changes in 
$L[{\bf n}]$: first, the typical energy scales of the system are slightly
renormalized, 
$J \rightarrow \tilde{J}$ and $\omega_0 \rightarrow \tilde{\omega}_0$,
even for $J^\prime = J $ and $k_z^\prime = k_z$. More importantly, $L$ acquires
an additional term due to the Zeeman energy and Berry phase of the 
uncompensated
spin $\delta s \, {\bf n}$. For all cases of experimental interest, due to
Eq.~(\ref{eq:3-assumpt2}) we have $h_A \ll \tilde{\omega}_0 \ll h_x$. 
The $h_A$-dependent terms in Eq.~(\ref{eq:3-lagrange}) hence lead only to 
minor modifications of the thermodynamic properties of the modified FW 
compared to the FW, but feature in the dynamics.

We will show now that, for $h_x \neq 0$, ${\bf n}$ and
hence the excess spin no longer trace a tunneling path in the $(y,z)$-plane. 
As can be seen from  
\begin{eqnarray}
&& L[\phi,\theta] \simeq \frac{N \hbar^2}{8 \tilde{J}} 
[-(h_x - i \dot{\phi})^2 -
\tilde{\omega}^2_0 \sin^2 \phi] - i \hbar \, \delta s \, \dot{\phi} \nonumber  \\
&& + \frac{N \hbar^2}{8 \tilde{J}} \Bigl[ 
\dot{\theta}^2 + \cos^2 \theta \left( (h_x - i \dot{\phi})^2 +
\tilde{\omega}^2_0 \sin^2 \phi \right) \nonumber \\ && + 2 \cos \theta \,
(h_x - i \dot{\phi}) \left(\delta s \, \frac{4 \tilde{J}}{N \hbar}
+ h_A \sin^2 \theta \sin^2 \phi \right)
\Bigr],
\label{eq:3-lagrangeb}
\end{eqnarray}
the timescales for the dynamics of
$\phi$ and $\theta$ separate due to Eq.~(\ref{eq:3-assumpt4}), and 
we can again invoke the adiabatic approximation used in 
Sec.~\ref{sec:0-dyn}. $\theta$ oscillates in a 
slowly varying harmonic potential with the potential minimum at 
$\theta_0$, where
\begin{equation}
\cos \theta_0 = - \frac{h_x - i \dot{\phi}}{(h_x - i \dot{\phi})^2 
+ \tilde{\omega}^2_0 \sin^2 \phi} \, \left( 
\frac{4 \tilde{J}}{N \hbar} \delta s + h_A \sin^2 \phi
\right).
\label{eq:3-thetamin}
\end{equation}
Corrections to the adiabatic approximation are beyond the order of the 
present calculation.
Eq.~(\ref{eq:3-thetamin}) shows that finite $\delta s \neq 0$ or 
$h_A \neq 0$ leads to a shift in the saddle-points of the Lagrangean $L[\phi,
\theta]$ away from $\theta_0=\pi/2$, which is due to the fact that then
${\bf S}_A - {\bf S}_B$ no longer lies in the $(y,z)$-plane 
[Fig.~\ref{fig3}c].
Expanding Eq.~(\ref{eq:3-lagrangeb}) to second order in $\vartheta= 
\theta - \theta_0$ and carrying out the Gaussian integral over $\vartheta$, 
we obtain a $\phi$-dependent effective Lagrangean 
\begin{eqnarray}
L[\phi] &= & \frac{N \hbar^2}{8 \tilde{J}}[-(h_x - i \dot{\phi})^2 
- \tilde{\omega}^2_0 \sin^2 \phi ] 
- i \hbar \left(\delta s + \frac{1}{2} \right) \dot{\phi} \nonumber \\
&& \hspace*{1cm}+ \hbar \frac{h_x}{2} + {\mathcal O}
(\tilde{\omega}_0^2/h_x, Nh_A \tilde{\omega}_0/8 \tilde{J}).
\label{eq:3-lagrangec}
\end{eqnarray}
Comparison with the corresponding Lagrangean for $\delta s=0$  
[Eq.~(\ref{eq:2-actionphi})] shows that, to leading order in the excess spin
$\delta s$ and anisotropy field $h_A$,
the only effect of an excess spin is to introduce an
additional topological phase $-i \delta s \, \dot{\phi}$. In particular, in 
contrast to the case $h_x=0$ discussed in earlier work on tunneling in 
ferrimagnets,~\cite{loss:92,duan:95,chudnovsky:95,chiolero:97} the potential 
barrier and hence
the real part of the tunnel action is only slightly altered by the excess 
spin. This is due to the fact~\cite{ivanov:99} that, for 
$\hbar h_x > |\delta s| 8J/N$ [Eq.~(\ref{eq:3-assumpt4})], the system is 
in the AF
regime in which the tunnel splitting is only slightly modified by the excess
spin.
Note that Eq.~(\ref{eq:3-lagrangec}) is formally identical to 
Eq.~(\ref{eq:2-actionphi}), which provides a rigorous proof of the statement 
that the modified FW also
may exhibit tunneling of ${\bf n}$ for sufficiently large anisotropy, as
already claimed on basis of physical arguments at the end of 
Sec.~\ref{sec:subl}.

\begin{figure}
\centerline{\psfig{file=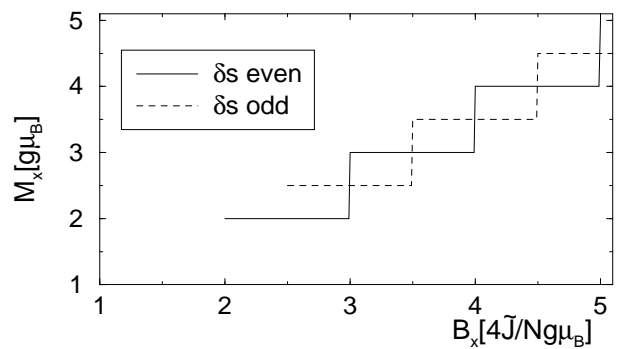,width=8.0cm}}
\caption{Schematic plot of the ground-state magnetization of a 
modified FW with integer (---)
and half-integer (- -) $\delta s$, respectively.}
\label{fig4}
\end{figure}

Using the same techniques as for the FW, we find
\begin{equation}
Z= \exp \left[\beta \left(\frac{N \hbar^2}{8 \tilde{J}} h_x^2  
-\hbar \frac{h_x+ \tilde{\omega}_0}{2}  
\right) \right] \, 
\cosh \left(\frac{\beta \tilde{\Delta}}{2}\right),
\label{eq:3-z}
\end{equation}
with the tunnel splitting 
\begin{equation}
\tilde{\Delta} (h_x) = \tilde{\Delta}_0 \left|\sin \pi
\left(\frac{N \hbar }{4 \tilde{J}} h_x + \delta s\right) 
 \right|,
\label{eq:3-delta}
\end{equation}
where $\tilde{\mathcal S}/
\hbar=N  \tilde{\omega}_0/2 \tilde{J}$, $\tilde{\Delta}_0 = 
8 \hbar \tilde{\omega}_0\sqrt{\tilde{\mathcal S}/2 \pi \hbar} 
\exp[-\tilde{\mathcal S}/\hbar]$. 
From Eq.~(\ref{eq:3-z}) it is also straightforward to derive all thermodynamic
quantities of interest. In particular, for the free energy $F$, the 
magnetization $M_x$, and the specific heat $c_v$ we obtain
\begin{eqnarray}
F&=& \hbar \frac{\tilde{\omega}_0+h_x}{2} - \frac{N \hbar^2}{8 \tilde{J}} 
h_x^2 - \frac{1}{\beta} \ln \cosh \left(
\frac{\beta \tilde{\Delta}}{2} \right), \label{eq:3-f} \\
M_x & = & (g \mu_B) \left[ \frac{N \hbar h_x}{4 \tilde{J}}-\frac{1}{2}  
+ \frac{1}{2 \hbar} \frac{\partial \tilde{\Delta}}{\partial h_x} 
\tanh \left( \frac{\beta \tilde{\Delta}}{2} \right)
\right] , 
\label{eq:3-mag}\\
c_V &=& k_B \left( \frac{\beta \tilde{\Delta}}{2} \right)^2 \cosh^{-2} 
\left( \frac{\beta \tilde{\Delta}}{2} \right)
\label{eq:3-cv}.
\end{eqnarray}
The most significant change in the thermodynamic properties of the 
modified FW is that, for half-integer 
$\delta s$, the zeros of $\tilde{\Delta}$ and hence the magnetization steps 
are shifted by a magnetic field $2 \tilde{J}/N g \mu_B$, i.e. half of a
magnetization plateau, compared to the 
unmodified wheel [Fig.~\ref{fig4}]. The magnetization plateaus then lie 
at half-integer spin values. At low $T$, the specific heat $c_v$ should
exhibit a characteristic Schottky anomaly, with a peak at $T_0 \simeq 0.4
\tilde{\Delta}/k_B$. So far, $c_v$ has been measured only 
for $B_x=0$ in Fe$_{10}$ and Fe$_6$ samples.~\cite{affronte:99} 
Measurements of $T_0$ in $c_v(T)$ for various 
$h_x \gg \tilde{\omega}_0$ would in principle allow one to observe the 
characteristic sinusoidal variation of $\tilde{\Delta}$ as function of $h_x$. 

\begin{figure}
\centerline{\psfig{file=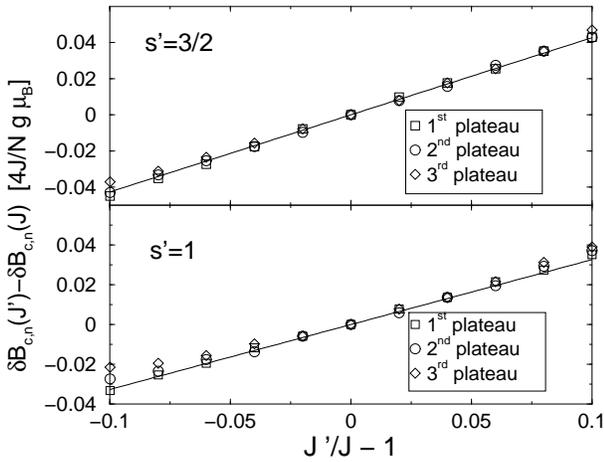,width=8.0cm}}
\caption{Comparison of ED (symbols) and analytical (---) 
results for the difference in plateau width, 
$\delta B_{c,n} (J^\prime) - \delta B_{c,n} (J^\prime = J)$ as 
function of $J^\prime-J$ for $\delta s = - 1/2$ 
(upper panel) and $\delta s = - 1$ (lower panel). $N=4$, $s=2$, $k_z =
k_z^\prime = 0.1 J$. The numerical error
of the data points $\pm 0.002 \times 4 J/N g \mu_B$ is smaller than the 
symbol size.}
\label{fig5}
\end{figure}

The width of the magnetization plateaus, 
$\delta B_{c,n} = B_{c,n+1}-B_{c,n} =  
4\tilde{J}/N g \mu_B$, where $B_{c,n}$ is the field at which the magnetization
$M_x$ exhibits the n$^{\rm th}$ step, is a quantity which is  
accessible in experiments and from which $J^\prime$ can be inferred.
Our theory predicts
\begin{equation}
\delta B_{c,n} (J^\prime) - \delta B_{c,n} (J^\prime = J) 
= 8 \frac{s^\prime /s}{(N + \delta s/s)^2} \frac{(J^\prime - J)}{g \mu_B}.
\label{eq:3-magvsjp}
\end{equation}
In Fig.~\ref{fig5} we compare the functional dependence predicted by 
Eq.~(\ref{eq:3-magvsjp}) with the results of
exact diagonalization (ED) on small rings, $N=4$, $s=2$, $s^\prime = 3/2$
(upper panel) and $s^\prime = 1$ (lower panel), respectively, for
$|J^\prime/J-1| \leq 0.1$.~\cite{remark2} The ED results are in good
agreement with the analytical result. The deviations for 
$|J^\prime/J-1| \gtrsim 0.1$
signal the breakdown of our ansatz in Eq.~(\ref{eq:3-staggering})  
for $J^\prime$ significantly different from $J$ (see 
Sec.~\ref{sec:TD-SD2} below).

We show now that although, for a given direction of ${\bf n}$, the total spin
${\bf S}$ acquires a component perpendicular to the field ${\bf B}$, the 
magnetization
$M_\alpha =0$ still vanishes for $\alpha=y,z$. We define the fields 
$m_\alpha (\tau)
= \delta \int {\rm d}\tau L[{\bf n}]/\delta (\hbar h_\alpha(\tau)) 
|_{h_\alpha \equiv 0}$ 
such that $M_\alpha = -(g \mu_B)
\int {\mathcal D} {\bf n} \, m_\alpha (\tau) e^{-\int {\rm d}\tau L/\hbar}/Z$. 
Using Eq.~(\ref{eq:3-lagrangeb}), we obtain
\begin{eqnarray}
m_z &=& \frac{N \hbar}{4 \tilde{J}} [-h_A n_z (1-n_z^2) \nonumber \\
&& \hspace*{2cm} + h_x n_x n_z + 
i ({\bf n} \times \dot{\bf n})_z]+ \delta s \, n_z \nonumber \\
& \simeq &  - \frac{N \hbar}{4 \tilde{J}} h_A \sin \phi \, + \delta s \,
\frac{\tilde{\omega}^2_0}{(h_x-i\dot{\phi})^2} \sin^3 \phi  \nonumber \\
&&  \hspace*{2cm}+ i \frac{N \hbar}{4 \tilde{J}} 
\dot{\theta_0} \cos \phi  + {\mathcal O}(\vartheta),
\label{eq:3-magfield}
\end{eqnarray}
with $\theta_0$ defined in Eq.~(\ref{eq:3-thetamin}).~\cite{remark2a} 
As follows from the invariance of $\hat{H}$ under rotation around 
${\bf B}$ by $\pi$, $M_y=M_z=0$ for
arbitrary $h_x$. In particular, at $T=0$, $M_z=0$ indicates that the 
ground-state is not a state with definite direction of ${\bf n}$, but rather a
coherent superposition of such states, $(|\uparrow\rangle + 
|\downarrow\rangle)/\sqrt{2}$, as expected for a system which shows coherent
N{\'e}el vector tunneling.

\section{Dynamics of the modified FW}
\label{sec:dyn}

As we have shown above, the effective action $L[\phi]$ of the
modified FW [Eq.~(\ref{eq:3-lagrangec})] is formally identical to that of 
the FW [Eq.~(\ref{eq:2-actionphi})]. In particular, for large anisotropy 
$k_z$, such that
$\tilde{\mathcal S}/\hbar \gg 1$, the modified FW is in the quantum tunneling 
regime [Sec.~\ref{sec:0-mqc}]. In this section, we evaluate explicitly the 
spin susceptibility $\chi_{zz}(\tau)$ for the modified FW. 

In order to motivate this, we first calculate $\chi_{zz}$ 
using the results of the classical vector model [Sec.~\ref{sec:subl}]. For 
${\bf n}$ along ${\bf e}_z$ or ${\bf -e}_z$, the $z$-component of the
total spin vector is finite, $S_z = 
\pm S_0$, where $S_0 = \delta s \, \omega_0^2/h_x^2 + 
(k^\prime_z s^{\prime 2} - k_z s^2)/2Js$. For ${\bf n}(t=0)={\bf e}_z$, the
coherent tunneling of ${\bf n}$ then results in an oscillating
$S_z(t) = S_0
\cos (\tilde{\Delta} t/\hbar)$, such that the Fourier transform of the 
(real-time) susceptibility
exhibits an absorption pole $\chi_{zz}^{\prime \prime}  
(\omega\simeq \tilde{\Delta}/\hbar) 
= \pi |S_0|^2 \tanh (\beta  \tilde{\Delta}/2) \delta(\omega- 
\tilde{\Delta}/\hbar)$.

Generalizing the procedure for the FW [Appendix~\ref{sec:a-suscept}] to 
$\delta s \neq 0$, we calculate the 
quantum corrections to this result from~\cite{allen:97}
\begin{eqnarray}
&& \chi_{zz}(\tau) = (g \mu_B)^2 \frac{N \hbar}{4\tilde{J}} 
(1-\langle \sin^2 \phi \rangle)
\delta(\tau ) \nonumber \\ && +  (g \mu_B)^2 \frac{1}{Z}
\int {\mathcal D}{\bf n}\, 
e^{-\int_0^{\beta \hbar} {\rm d} \tau L[{\bf n}]/\hbar} m_z(\tau) 
m_z(0),
\label{eq:4-imagtimesusc}
\end{eqnarray}
with $m_z(\tau)$ given in Eq.~(\ref{eq:3-magfield}). As for the undoped 
FW, the correlations of the $\vartheta$-terms in $m_z(\tau)$ give rise to a 
strongly peaked term $ (g \mu_B)^2 N \hbar \delta(\tau)\langle \sin^2 \phi 
\rangle /4 \tilde{J} $. We hence find~\cite{remark3}
\begin{eqnarray}
&& \chi_{zz}(\tau) = (g \mu_B)^2   \frac{N \hbar}{4\tilde{J}} 
\delta(\tau ) \label{eq:4-imagtimesuscb}
 \\ &&+ (g \mu_B)^2  \frac{1}{Z}
\int {\mathcal D}\phi \, 
e^{-\int_0^{\beta \hbar} {\rm d} \tau L[\phi]/\hbar} 
\left(-\frac{N \hbar h_A}{4 \tilde{J}} \sin \phi (\tau) \right.  
\nonumber \\ && \left.
+ \delta s \,
\frac{\tilde{\omega}^2_0}{h_x^2} \sin^3 \phi (\tau) \right)
\left(-\frac{N \hbar h_A}{4 \tilde{J}} 
\sin \phi+ \delta s \,
\frac{\tilde{\omega}^2_0}{h_x^2} \sin^3 \phi \right).\nonumber
\end{eqnarray}
In stark contrast to the FW, the path integral in 
Eq.~(\ref{eq:4-imagtimesuscb}) gives rise to terms proportional to
$\exp[\pm \tilde{\Delta} |\tau|/\hbar ]$ such that, upon Fourier 
transform, 
the susceptibility in Matsubara representation contains terms
$1/(i \omega_n \pm \tilde{\Delta}/\hbar)$. The path integral is most  easily
evaluated in a Hamiltonian description. We requantize the field $\phi$ and 
use an effective two-state Hamiltonian to evaluate the matrix elements.
Inserting the expression for $h_A$, we find [Appendix~\ref{sec:a-TSS}]
\begin{eqnarray}
&& |\langle e| \hat{S}_z|g \rangle| = \Bigl| \frac{N (k_z^\prime s^{\prime 2}
- k_z s^2)}{2 \tilde{J} 
s(N+\delta s/s)} \left( 1-\frac{\tilde{J}}{N \hbar \tilde{\omega}_0} 
\right)  \nonumber \\ && \hspace*{2cm}+ \delta s \,
\frac{\tilde{\omega}^2_0}{h_x^2} \left( 1-3 \frac{\tilde{J}}{N \hbar
\tilde{\omega}_0} \right) \Bigr|, \label{eq:4-matrixel} \\
&& \chi^{\prime \prime}_{zz} (\omega \simeq \tilde{\Delta}) = \pi (g \mu_B)^2 
|\langle e|\hat{S}_z|g \rangle|^2 \,\nonumber \\ && \hspace*{2cm} \times 
\tanh \left( \frac{\beta \tilde{\Delta}}{2} \right) \,
\delta(\omega - \tilde{\Delta}/\hbar).
\label{eq:4-imagtimesuscd}
\end{eqnarray}
Eqs.~(\ref{eq:4-matrixel}) and (\ref{eq:4-imagtimesuscd}) are the main results 
of this section. For $\delta s = -5/2$ or $-1$ (for Ga and Cr dopants, 
respectively), $|\langle e|\hat{S}_z|g \rangle|$ can 
be of  order $0.1$ even for $h_x \gtrsim 3 \tilde{\omega}_0$. For 
$k_B T \lesssim \tilde{\Delta}$ 
the susceptibility of the modified FW then exhibits a resonance 
at $\omega=\pm \tilde{\Delta}/\hbar$ which is accessible in AC susceptibility
or ESR measurements. The
terms $\tilde{J}/N  \hbar \tilde{\omega}_0=1/(2 \tilde{\mathcal S}/\hbar)$ 
in $|\langle e|S_z|g \rangle|$
are quantum corrections to the classical result derived at the beginning of 
this section.

So far we have ignored decoherence of the spin tunneling, which is
crucial for the notion of MQC. The condition 
$\Gamma \simeq \tilde{\Delta}/\hbar$, where $\Gamma$ is the electron spin 
decoherence rate, 
marks the  transition from coherent to incoherent tunneling dynamics. As is 
evident 
from the classical vector model discussed in Sec.~\ref{sec:subl}, $S_z$ 
follows 
the tunneling dynamics of ${\bf n}$. In particular, for a single modified FW, 
the 
decay rate of $|\langle n_z (t) n_z \rangle|$, $\Gamma$, is also the decay 
rate of $|\langle S_z (t) S_z \rangle|$. For $\Gamma \neq 0$, the 
$\delta$-peak in Eq.~(\ref{eq:4-imagtimesuscd}) is then broadened into a 
Lorentzian of width $\Gamma$. 
In experiments carried out on an ensemble of modified FWs, inhomogeneous 
broadening (e.g. due to crystal defects or nuclear spins) adds to the width 
of the resonance peaks. The experimentally determined linewidth of the 
absorption and
emission peaks provides an upper limit for $\Gamma$.
This should allow one to settle the experimentally  
unresolved problem of
whether true quantum coherence is established in ferric wheels.

\begin{figure}
\centerline{\psfig{file=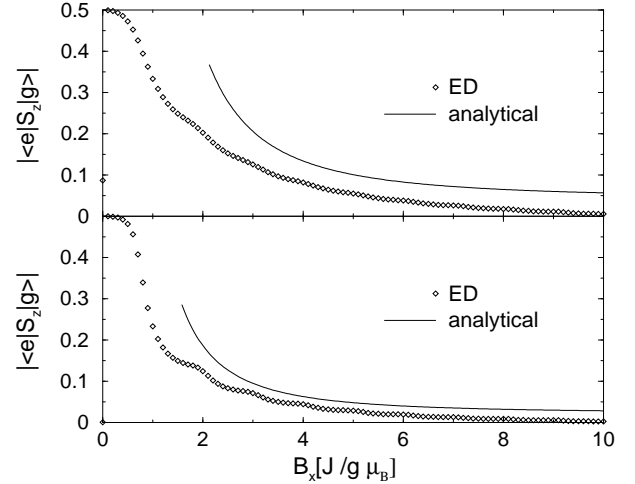,width=8.0cm}}
\caption{Transition matrix element $|\langle e|S_z|g\rangle|$ for small rings,
$N=4$ with $J^\prime = J$, $s=5/2$, $s^\prime = 2$. 
In the upper panel, $k_z = k_z^\prime= 0.1$.
In the lower panel, $k_z=k_z^\prime = 0.055 J$ is chosen such that 
$\tilde{{\mathcal S}}/\hbar \simeq 3.3$ as for Fe$_{10}$. The analytical
result (---) is shown for $\max[\hbar \tilde{\omega}_0,
|\delta s|8J/N] < \hbar h_x < 4 J s$. Due to Eq.~(\ref{eq:3-assumpt4}), our
theory is rigorously valid only for fields much larger than 
$\max[\hbar \tilde{\omega}_0,|\delta s|8J/N]$ and much smaller than $4Js$.
}
\label{fig6}
\end{figure}

\begin{figure}
\centerline{\psfig{file=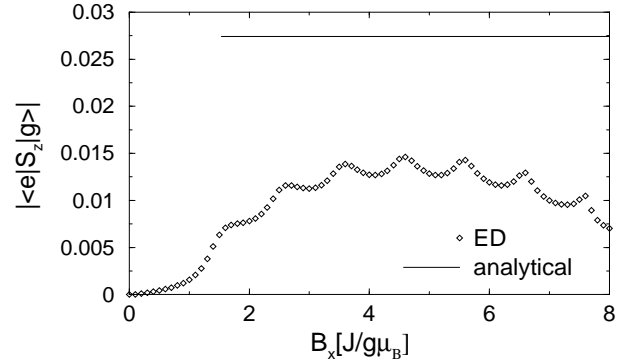,width=8.0cm}}
\caption{Transition matrix element $|\langle e|S_z|g\rangle|$ for a small ring 
$N=4$ with $J^\prime=J$, $s=s^\prime =2$, i.e. $\delta s =0$, but 
$k_z^\prime = 1.5 k_z$ and hence $h_A\neq 0$. 
Again, $k_z=0.0655 J$ is chosen such that 
$\tilde{{\mathcal S}}/\hbar \simeq 3.3$, as for Fe$_{10}$. The analytical
result (---) is shown for $\max[\hbar \tilde{\omega}_0,
|\delta s|8J/N] < \hbar h_x < 4 J s$. See also caption of Fig.~\ref{fig6}.
}
\label{fig7}
\end{figure}

\begin{figure}
\centerline{\psfig{file=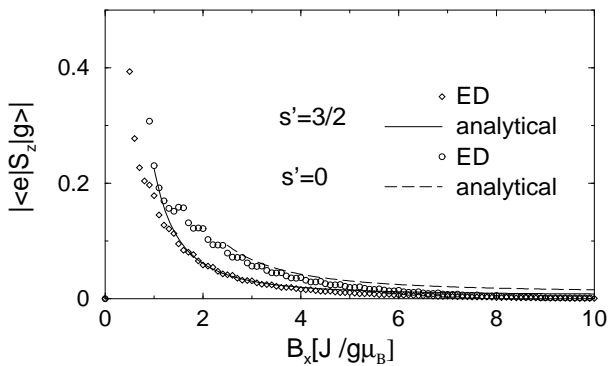,width=8.0cm}}
\caption{Transition matrix element $|\langle e|S_z|g\rangle|$ obtained with the
phenomenological sublattice Hamiltonian [Eq.~(\ref{eq:3-sublh})] for  
Fe$_{10}$ with
one $s=5/2$ substituted by (a) a dopant with $s^\prime=3/2$ (e.g. Cr) 
(numerical data
$\diamond$, analytical prediction ---),  
and (b) a dopant with $s^\prime=0$ (e.g. Ga) (numerical data
$\circ$, analytical prediction - -). Note that, in this case, the numerical 
data
is not obtained from ED of Eq.~(\ref{eq:1-h}), but rather
of Eq.~(\ref{eq:3-sublh}). For simplicity, 
we assumed $J^\prime=J$ and $k^\prime_z = k_z = 0.0088 J$. 
The analytical
results (--- and - -) are shown for $\max[\hbar \tilde{\omega}_0,
|\delta s|8J/N] < \hbar h_x < 4 J s$. See also caption of Fig.~\ref{fig6}.}
\label{fig8}
\end{figure}

Finally we compare our result for the transition matrix element $|\langle
e | S_z |g \rangle |$ entering Eq.~(\ref{eq:4-imagtimesuscd}) with results 
obtained from ED on small rings for a wide range of parameters 
[Figs.~\ref{fig6}, \ref{fig7}, \ref{fig8}]. For simplicity, we assume 
$J^\prime = J$. In the range of validity of our theory 
[Eq.~(\ref{eq:3-assumpt4})], for $\delta s\neq 0$ [Figs.~\ref{fig6}, 
\ref{fig8}], the agreement
of ED ($\diamond$) with analytic results (---) is both qualitatively and
quantitatively convincing.~\cite{remark3a} The 
small oscillating 
features seen in the exact results are due to tunneling corrections ${\mathcal 
O}(
\exp [-{\tilde{\mathcal  S}}/\hbar])$ to  $|\langle e | S_z |g \rangle |$, 
which were neglected in Eq.~(\ref{eq:4-matrixel}). 
For Fig.~\ref{fig7}, where $\delta s = 0$, our theory makes the correct 
qualitative prediction that $|\langle e | S_z |g \rangle |$ depends only
weakly on $h_x$, but overestimates the matrix element. However, due to the 
smallness of the matrix element for $\delta s=0$, the discrepancy can be 
due to terms neglected in the derivation of Eq.~(\ref{eq:3-lagrangec}).
The significantly
different qualitative features of Fig.~\ref{fig7} compared to Figs.~\ref{fig6}
and \ref{fig8} arise from the fact that $\delta s=0$ in Fig.~\ref{fig7}. The 
different functional dependence of $|\langle e | S_z |g \rangle |$ on $h_x$ 
for the two cases $\delta s \neq 0$ and $\delta s = 0$ is 
well understood within the theoretical framework presented here 
[Eq.~(\ref{eq:4-matrixel})]. The very large difference in matrix element
magnitude illustrates the importance of looking at doped rings in 
experiment.

We conclude this section by remarking that, for finite excess spin 
$\delta s$, the second transverse susceptibility 
$\chi^{\prime \prime}_{yy} (\omega)$ also has an absorption pole at $\omega = 
\tilde{\Delta}/\hbar$. However, since ${\bf e}_y$ is a hard axis, 
the spectral weight of this pole is 
significantly smaller than that of $\chi^{\prime \prime}_{zz} (\omega)$, such 
that N{\'e}el vector tunneling in the modified FWs can be more easily
detected by probing the latter quantity.

\section{Thermodynamics and Spin Dynamics for $J^\prime/J\gg 1$ and 
$J^\prime/J \ll 1$}
\label{sec:TD-SD2}

\begin{figure}
\centerline{\psfig{file=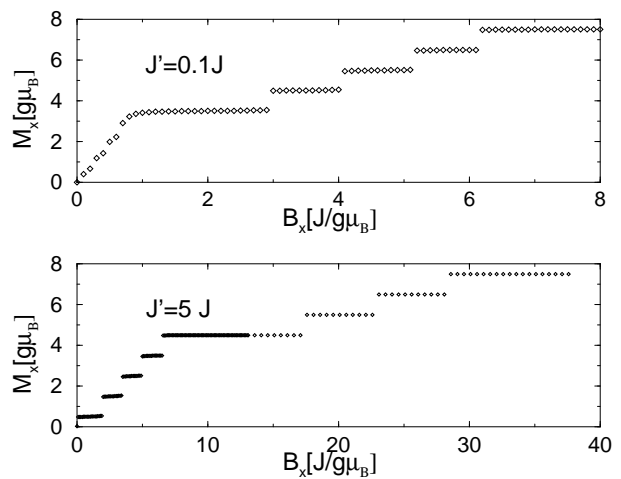,width=8.0cm}}
\caption{$M_x(B_x)$ at $T=0$ for a small system with 
$J^\prime \ll J$ (upper panel) or
$J^\prime \gg J$ (lower panel). Here, $N=4$, $s=2$, $s^\prime=3/2$,
$k_z=k_z^\prime=0.1$.}
\label{fig9}
\end{figure}

The deviations of the ED results from our theoretical predictions shown in 
Fig.~\ref{fig5} indicate that, for $J^\prime/J \ll 1$ or 
$J^\prime/J \gg 1$, 
the theory developed in Sec.~\ref{sec:thermodyn} is no longer 
immediately applicable. Indeed, results obtained by ED for the ground-state 
magnetization $M_x$ in small rings ($N=4$, $s=2$, and $s^\prime = 3/2$)
[Fig.~\ref{fig9}] indicate that one of the main results of 
Sec.~\ref{sec:thermodyn}, that  $M_x$ exhibits a series of equally 
spaced magnetization steps, does not hold any more. As we will show below,
this is due to the fact that our ansatz Eq.~(\ref{eq:3-staggering}) needs
to be modified for $J^\prime$ significantly different from $J$. 
In this section we show that, for the limiting cases of $J^\prime 
\gg J$ or $J^\prime \ll J$, the modified FW can be mapped onto the problem 
discussed in the preceding sections. We discuss the qualitative 
features of $M_x$ for these systems and show that coherent tunneling of 
${\bf n}$ also results in coherent oscillations of the total spin.

${\mathit J^\prime \ll J:}$ In this limit,  
${\bf s}_1$ decouples from all other spins and aligns antiparallel to 
${\bf B}$ for $\hbar h_x \gtrsim J^\prime s$. The
remaining spins ${\bf s}_2$, ${\bf s}_3$, \ldots ${\bf s}_N$ form an open
spin chain, as sketched in Fig.~\ref{fig10}a. As shown
in Appendix~\ref{sec:a-openchain}, the Lagrangean of an open spin chain with 
an odd number of spins can also be mapped onto Eq.~(\ref{eq:3-lagrangec}), 
with $\delta s = s$ and slightly renormalized 
$\tilde{J}=JN(N-2)/(N-1)^2$. We predict that $M_x$ has the following 
features:
\begin{itemize}
\item $M_x \gtrsim g \mu_B (s^\prime + s)$ for $2J^\prime s \ll \hbar h_x$.
\item For ${\rm max}[\hbar \tilde{\omega}_0, s 8 \tilde{J}/N] \ll \hbar h_x 
\ll 4 \tilde{J} s$, $M_x$   exhibits a series of equally spaced 
magnetization steps with plateau width $\delta B_{c,n}= 4 \tilde{J}/N 
g \mu_B$. Depending on whether
$\delta s$ is half-integer or integer, the plateaus  
correspond to states with half-integer or integer total spin, respectively.
\end{itemize}

${\mathit J^\prime \gg J:}$ In this limit, the spins ${\bf s}_N$, ${\bf s}_1$ 
and ${\bf s}_2$ are strongly coupled. In a semiclassical picture, 
${\bf s}_1$ aligns antiparallel 
to ${\bf s}_N$ and ${\bf s}_2$ and the three spins act as one single spin 
$|2s - s^\prime|$ coupled to ${\bf s}_3$ and 
${\bf s}_{N-2}$ with exchange constant $J$ [Fig.~\ref{fig10}b].
For simplicity we assume 
$s^\prime<2s$. Then, for $\hbar h_x \ll  J^\prime  (s-\delta s+1)$, $\hat{H}$ 
can be mapped onto a Hamiltonian of the
 form Eq.~(\ref{eq:1-h}) with $N \rightarrow N-2$, $J^\prime \rightarrow J$, 
$k_z^\prime \rightarrow (k_z^\prime s^{\prime 2} + 2 k_z s^2)/
(2s-s^\prime)^2$, and $s^\prime \rightarrow 2s - s^\prime = s - \delta s$. 
Because all $N-2$ exchange couplings in the new Hamiltonian are identical, 
the theory developed in Sec.~\ref{sec:thermodyn} and 
\ref{sec:dyn} remains applicable. In particular, for the ground-state
magnetization $M_x$ we make the following predictions:
\begin{itemize}
\item For ${\rm max}[\hbar \tilde{\omega}_0, |\delta s|8J/(N-2)] \ll \hbar h_x 
\ll 4 J s$, $M_x$ exhibits a series of equally spaced magnetization steps
with $\delta B_{c,n}\simeq 4J/(N-2) g \mu_B$. Depending on whether 
$\delta s$ is
half-integer or  integer, the plateaus correspond to states with 
half-integer or integer total spin, respectively.
\item For $\hbar h_x \gtrsim J^\prime (s-\delta s+1)$, the Zeeman energy is 
sufficiently large to destroy the AF configuration of ${\bf s}_N$, 
${\bf s}_1$, and ${\bf s}_2$. This results in a series of additional 
magnetization steps with spacing $J^\prime$. 
\end{itemize}

Note that a similar argument also applies if $J^\prime < 0$. In this case, the
three spins ${\bf s}_N$, ${\bf s}_1$, and ${\bf s}_2$ are ferromagnetically 
coupled and align parallel. Again, the system can be mapped onto a smaller
ring (as in Fig.~\ref{fig10}b), where now $s^\prime \rightarrow 2 s + 
s^\prime$. The magnetization curve resembles the one shown in the upper
panel of Fig.~\ref{fig9}.
In Fig.~\ref{fig9}, ED results for small rings with
$N=4$, $s=2$, $s^\prime=3/2$ are displayed. The qualitative features 
agree with all the above predictions.

\begin{figure}
\centerline{\psfig{file=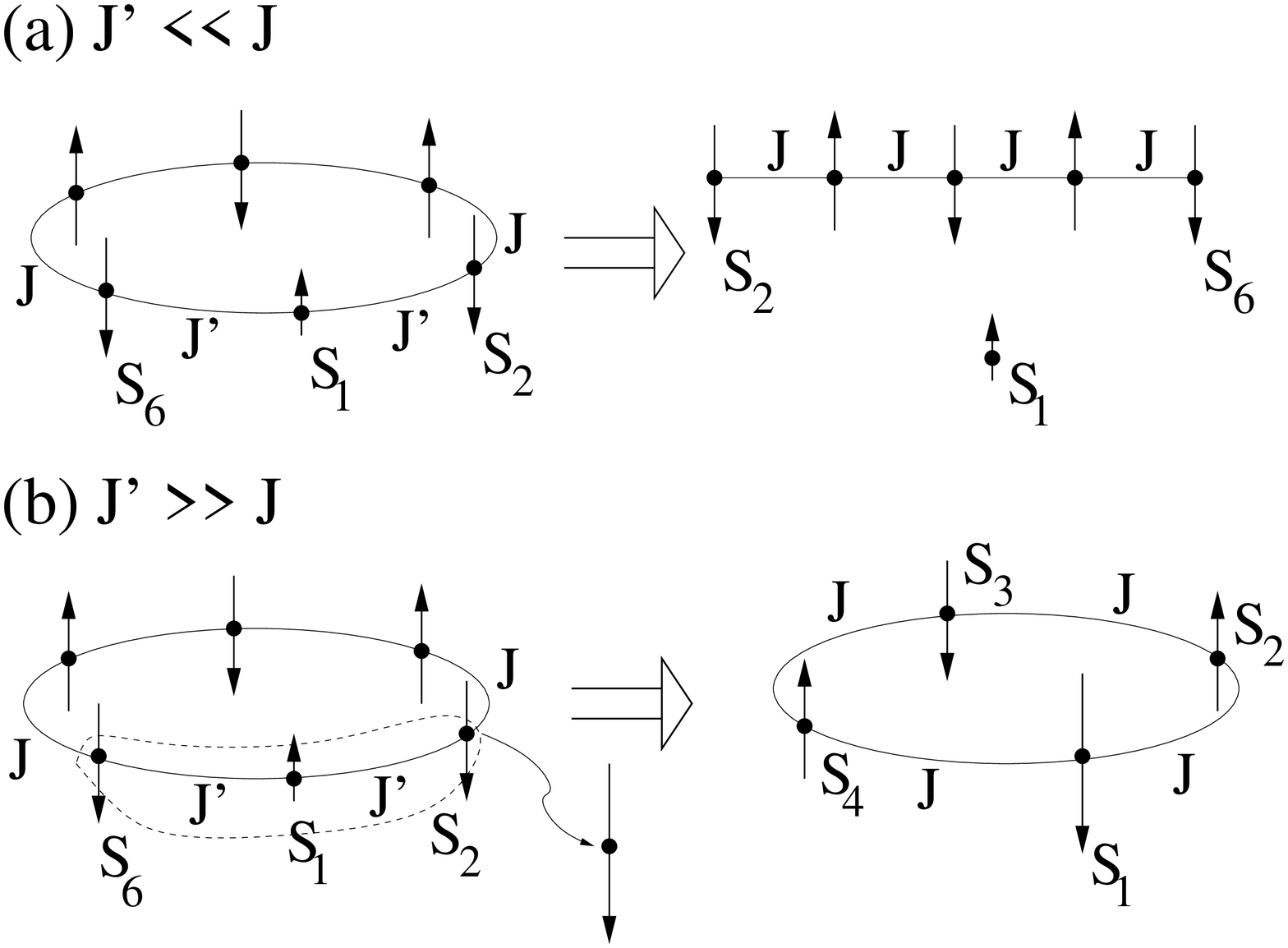,width=8.0cm}}
\caption{(a) For $J^\prime \ll J$, ${\bf s}_1$ decouples from all other spins
and the modified FW can be mapped onto an open spin chain excluding 
${\bf s}_1$. (b) For $J^\prime \gg J$, ${\bf s}_N$, ${\bf s}_1$, and 
${\bf s}_2$ are strongly coupled such that they can be described as one single
large spin. The Hamiltonian of the ring then maps onto that of a modified FW
with $J^\prime=J$.}
\label{fig10}
\end{figure}

We conclude that even the qualitative features of $M_x$ allow one to estimate
the parameter $J^\prime$ of a modified FW. Even more important, 
as we 
have shown, also for $J^\prime \ll J$ and $J^\prime \gg J$ the modified FW can
be mapped onto the Lagrangean [Eq.~(\ref{eq:3-lagrange})] of 
a system which exhibits quantum tunneling
of ${\bf n}$. In all cases discussed above, the
quantum tunneling of ${\bf n}$ leads to coherent oscillations of 
the total spin ${\bf S}$, and thus can be observed in AC susceptibility or
ESR measurements.

\section{Discussion}
\label{sec:discussion}

The theory described in Secs.~\ref{sec:thermodyn}, \ref{sec:dyn},
and \ref{sec:TD-SD2} allowed us to derive explicit expressions for both
thermodynamic quantities [Eqs.~(\ref{eq:3-mag}) and (\ref{eq:3-cv})] and the
susceptibilitiy $\chi_{zz}^{\prime \prime}$ [Eqs.~(\ref{eq:4-matrixel}) and 
(\ref{eq:4-imagtimesuscd})] of modified FWs. In order to establish a 
connection with experimental issues, we now outline the steps 
necessary to detect 
coherent N{\'e}el vector tunneling.
For simplicity we restrict our considerations to Fe$_{10}$ with
Ga ($\delta s = -5/2$) or Cr ($\delta s = -1$) impurity ions, and assume
$J^\prime \simeq J$.

For finite excess spin $\delta s$, the two energetically degenerate spin
configurations [Fig.~\ref{fig1}] required for coherent spin tunneling as
discussed in the present work 
certainly exist if $\hbar h_x \gg |\delta s|8J/N$ [Eq.~(\ref{eq:3-assumpt4})]. 
This tunneling
regime is well within experimental reach for Cr dopants
($B_x \gg 9$T), but not for Ga dopants ($B_x \gg 23$T).~\cite{remark4}
For Cr  dopants ($s^\prime \neq 0$), however, the two new parameters 
$J^\prime$ and $k_z^\prime$ introduced in $\hat{H}$ [Eq.~(\ref{eq:1-h})] 
must first be determined in order to characterize the ring system.

Both $J^\prime$ and $k_z^\prime$ can be obtained from the measurement of
two independent thermodynamic quantities, such as the ground-state
magnetization and tunnel splitting. A schematic plot of the ground-state 
magnetization for integer $\delta s$ ($\circ$) is shown in 
Fig.~\ref{fig4}.
Although the magnetization steps are smeared out at finite temperature, for
$T \lesssim 1$K,
the magnetization step spacing $\delta B_{c,n}$ still can be obtained 
with high 
accuracy.~\cite{taft:94} With $\delta B_{c,n} = 4 \tilde{J}/N g \mu_B$
and Eq.~(\ref{eq:3-newparameters}), this allows one to determine $J^\prime$.
The on-site anisotropy $k_z^\prime$ can be obtained from $\tilde{\Delta}$,
which   depends sensitively
on the tunnel action $\tilde{S} \propto \sqrt{(N-1) k_z s^2 +
k_z^\prime s^{\prime 2} }$ and hence on $k_z^\prime$. The tunnel splitting 
$\tilde{\Delta}$ (and hence $k_z^\prime$) is accessible either in AC 
susceptibility or ESR
measurements [Eq.~(\ref{eq:4-imagtimesuscd})], or in measurements of
thermodynamic quantities, such as $c_v$. Torque magnetometry
is another experimental technique which has been used
to determine the anisotropy constant with quasi-spectroscopic 
accuracy.~\cite{cornia:99}

Once $J^\prime$ and $k^\prime_z$ are known, 
Eq.~(\ref{eq:4-imagtimesuscd}) determines both the position and the 
spectral weight of the resonance in $\chi^{\prime \prime}_{zz} (\omega)$ 
which arises from coherent
quantum tunneling of ${\bf n}$. The characteristic functional dependence
of $\tilde{\Delta}$ [Eq.~(\ref{eq:3-delta})] and 
$|\langle e|S_z|g\rangle|^2$ [Eq.~(\ref{eq:4-matrixel})] on $h_x$ 
predicted by our theory can be checked experimentally. Finally, it
is noteworthy that, although $\tilde{\Delta}$ can be determined from 
thermodynamic quantities, the key problem of MQC is the
measurement of the decoherence rate $\Gamma$ which is accessible only in
dynamic quantities, such as AC susceptibilities.

Throughout the current work, we have considered FWs with only one
dopant ion. As we have shown in the preceding sections, thermodynamic
and dynamic quantities of doped FWs may differ significantly from those
of undoped FWs. In the large samples investigated experimentally, doping
will lead to a random distribution both of the number of dopant ions 
and of their position relative to the direction of the 
magnetic field. 
We defer a detailed analysis of these issues to a future publication. 
Here we note only that the random distribution of impurities does not 
invalidate the above considerations, and stress the qualitative
features which ensure this. The choice of a low impurity concentration 
results in a large majority of the FWs containing no dopants or having 
only one dopant ion, which allows one to extract the 
system parameters of the singly doped FWs. When
intraring dipolar interactions make a significant contribution to the 
effective uniaxial anisotropy $k_z$, doping with only
one ion changes the effective anisotropy from uniaxial to biaxial.
The theoretical framework presented in this paper can be readily extended 
to account for biaxial anisotropies. Because the original 
uniaxial anisotropy dominates the biaxial correction, the altered 
tunnel splittings $\tilde{\Delta}_i$ in a singly doped FW have a 
magnitude similar to $\tilde{\Delta}$ for the undoped FW, and a 
separation which is small by comparison. Thus AC susceptibility or 
ESR measurements can be expected to observe signals corresponding to
reversals of the total spin accompanying N{\'e}el vector tunneling, and 
governed by the frequencies $\tilde{\Delta}_i$.

\section{Spin Quantum tunneling in Ferritin}
\label{sec:ferritin}
The theoretical framework developed in this work is quite general and applies
to other systems besides AF ring systems. In particular, the results
of the classical sublattice model [Sec.~\ref{sec:subl}] can be easily
extended to different systems. In order to illustrate this point,
we now discuss natural horse-spleen ferritin and artificial magnetoferritin, 
in which spin quantum tunneling has already
been studied experimentally~\cite{awschalom:92,awschalom:92b,gider:95} and
theoretically.~\cite{duan:95,chudnovsky:95,chiolero:97,ivanov:99,harris:99} The
experiments were carried out in the presence of small static magnetic fields
($B_x \lesssim 10^{-6}$T). In this regime, $B_x \neq 0$ leads to an 
energy bias between the states $|\uparrow\rangle$ and 
$|\downarrow\rangle$, and tunneling is suppressed for increasing $B_x$. 

Recently it was shown~\cite{ivanov:99} 
that, for sufficiently large field, i.e. 
$\hbar h_x \sim |\delta s| \, 8J/N$, there are again
two energetically degenerate spin configurations between which spin quantum 
tunneling may take place. In natural horse-spleen ferritin,~\cite{harris:99} 
$J \simeq 200$K and $\delta s/N \simeq 0.05$. For a system with uniaxial 
hard axis anisotropy, $\hat{H}_{an,z} = k_z \sum_i \hat{s}_{i,z}^2$, the 
tunnel barrier and hence the tunnel action can be effectively controlled
over a wide range of parameters by varying the magnetic field 
$B_x$ applied in the easy plane.~\cite{ivanov:99} Tunneling
in the plane perpendicular to ${\bf B}$  gives rise to a
topological phase acquired by the 
spins.~\cite{garg:93,golyshev:95,wernsdorfer:99}
A drawback of the setup considered in Ref.~\onlinecite{ivanov:99} is that,
for uniaxial hard axis anisotropy, spin tunneling leaves the total spin 
${\bf S}$ invariant if ${\bf B}$ is applied perpendicular to the hard axis. 
As for the FW with equal sublattice spins, spin tunneling cannot be 
observed in AC susceptibility measurements.

However, experiments indicate that, in addition to the strong hard-axis 
anisotropy, ferritin also exhibits a second weak hard-axis anisotropy 
$\hat{H}_{an,y} = k_y \sum_i \hat{s}_{i,y}^2$, where~\cite{harris:99} 
$k_y/k_z \simeq 10^{-3}$. In self-sustaining films of natural horse-spleen
ferritin, the hard axis ${\bf e}_z$ is perpendicular to the 
film.~\cite{gider:95} In the simplest experimental setup, interference 
of different spin tunnel paths then could be explored if a field ${\bf B}=B_z
{\bf e}_z$ is applied along the hard axis. As long as 
$\hbar h_z \ll N k_z s^2/\delta s$ 
($\simeq 10$K for horse-spleen ferritin), due to the large anisotropy energy 
the spins remain confined to
the film plane such that there are again two energetically degenerate spin 
configurations similar to Fig.~\ref{fig1}. Tunneling takes place in 
the plane perpendicular to ${\bf B}$, with a tunnel splitting
\begin{equation}
\Delta = \Delta_0  \left|\cos \pi\left(S_{\rm tot} + \frac{N \hbar}{4 J} h_z 
\right) \right|,
\label{eq:ferritin-Delta}
\end{equation} 
where $\Delta_0 \simeq 5\times 10^{-5}$K and the total staggered spin is
$S_{\rm tot} \simeq 2.5 \times 4500$ for natural horse-spleen ferritin.
$\Delta$ is periodic as a function of $B_z$, with a period
$\delta B_z =  4 J/N g \mu_B$ ($\simeq 0.13$T for
natural horse-spleen ferritin). The advantage of this tunnel scenario is
that
quantum tunneling of ${\bf n}$ also results in a tunneling of the excess
spin $\delta s$, and hence leads to a large resonance peak in the
susceptibility per ferritin molecule, i.e.
\begin{equation}
\chi^{\prime \prime}_{xx} (\omega \simeq \Delta/\hbar) 
\simeq \frac{1}{2}  \pi  (\delta s)^2 \, \tanh
\left( \frac{\beta \Delta}{2} \right) \, \delta (\omega - \Delta/\hbar).
\label{eq:ferritin-susc}
\end{equation}
The factor $1/2$ takes into account the random distribution of easy axes
in the film plane. Due to the spread in particle number, the total staggered
spin of the system can be either integer or half-integer. Hence, one will
observe two different tunnel splittings varying with $h_z$ as 
$\Delta = \Delta_0 |\cos (\pi N \hbar h_z/4 J) |$ and $\Delta = \Delta_0 
|\sin (\pi N \hbar h_z/4 J) |$. An experimental confirmation of this behavior 
would provide further strong evidence
that the resonance observed in AC susceptibility measurements in 
ferritin~\cite{awschalom:92,awschalom:92b,gider:95} is due to macroscopic
spin tunneling. In particular, the period of the oscillations of $\Delta$
as function of the applied field $B_z$ would allow one to estimate the total
number of tunneling spins.

\section{Conclusion}
\label{sec:conclusions}

The AF ring systems discussed here, modified ferric wheels
which are already available to experimentalists, combine the advantages of
AF and ferromagnetic molecular magnets. The tunnel splitting $\tilde{\Delta}$ 
is sufficiently large
that quantum coherence between macroscopically different states is
established. Tunneling of the N{\'e}el vector ${\bf n}$ also leads to a 
tunneling of the total spin ${\bf S}$, thus making the spin dynamics 
in modified FWs accessible to experiment.
We have considered the simplest realistic model Hamiltonian $\hat H$
[Eq.~(\ref{eq:1-h})] for a system in which, for example,
one of the Fe ions is exchanged by Cr or Ga. We showed that the 
additional parameters entering $\hat{H}$ can be inferred from 
equilibrium quantities such as the magnetization. Moreover, for a wide
range of parameters, the system still exhibits MQC in the form of coherent
tunneling of ${\bf n}$. 
Finally, we calculated spin correlation functions of the modified FW and 
showed that tunneling of ${\bf n}$ can indeed be observed in AC 
susceptibility 
or ESR experiments, which allow one to measure both the tunnel
splitting $\tilde{\Delta}$ and an upper bound for the spin decoherence 
rate $\Gamma$.
Hence they should be appropriate to verify experimentally that N{\'e}el vector 
tunneling in FWs is coherent.        

Throughout this work we have used spin coherent-state path integrals leading
to a NLSM-description for the modified FW. The main advantage of this 
technique over ED is that thermodynamic and dynamic quantities can be
evaluated for a realistic system size. In addition, an intuitive physical 
understanding of the spin dynamics (quantum tunneling of ${\bf n}$) can 
be obtained. A drawback of the analytical approach chosen in the present
work is that it naturally requires approximations. Corrections to our results,
in particular $1/s$-corrections, may become appreciable 
for the parameters of the FWs. However, our analytical 
results for $|\langle e| \hat{S}_z |g \rangle|$ agree well with ED results 
obtained for small systems. For the parameter range explored in ED, deviations
from our theoretical predictions become significant mainly if $s^\prime$ 
is small ($s^\prime=1/2$ or $1$), where the ansatz Eq.~(\ref{eq:3-staggering}) 
fails due to large quantum fluctuations of $\hat{\bf s}_1$.
Although numerical work on rings with $N=6$, $8$, and $10$ is challenging, 
some 
results have been obtained for the ground-state magnetization and 
torque.~\cite{normand:00} A detailed numerical study also of
spin correlation functions would provide clearer
evidence for the range of validity of the present approach and could 
explore its limitations. For example, as is well 
known, the sublattice Hamiltonian Eq.~(\ref{eq:2-sublham})
is exact for $N=4$. For increasing $N$, however, ED results~\cite{schnack:00} 
show that there are small deviations from $\hat{H}_{0,{\rm subl}}$. 

Note that our work also has important implications for undoped ferric wheels.
Recent torque,~\cite{cornia:pc} $c_v$,~\cite{cornia:pc} and proton 
$1/T_1$-measurements~\cite{lascialfari:pc} on single crystals of various Fe$_6$
compounds indicate that these FWs could exhibit physics beyond
the Hamiltonian Eq.~(\ref{eq:1-h0}). One important future step in
explaining the new experimental data will be to clarify to which extent the
observed phenomena can be attributed to inhomogeneous level broadening. The 
theoretical framework presented here allows one to calculate 
analytically the inhomogeneous level broadening resulting from a random 
distribution of single exchange couplings $J^\prime$ and on-site anisotropies
$k_z^\prime$ which could be a consequence of lattice defects
in Fe$_6$ crystals. Indeed, recent work on Mn$_{12}$ suggests that
lattice distortions~\cite{chudnovsky:01} and a distribution of 
anisotropy energies and $g$-factors~\cite{park:01,mertes:01} could 
account for the 
observed broad distribution of tunneling rates in  Mn$_{12}$.

Finally, we stress once more that the Hamiltonian $\hat{H}$ 
[Eq.~(\ref{eq:1-h})]
discussed in this paper is a simple model Hamiltonian, which still leads
to fascinating novel features in the physical properties of the modified FWs.
However, as discussed in Sec.~\ref{sec:discussion},
realistic systems might require modification of Eq.~(\ref{eq:1-h}). 
Generalization of the present 
approach to more complicated anisotropies is, however, straightforward.

\acknowledgments
This work was supported in part by the European Network 
MolNanoMag under grant number HPRN-CT-1999-00012, by the Federal Office for
Education and Science (BBW) Bern, and by the Swiss National Fund. We are 
indebted to F.~Borsa, A.~Cornia, V.~Golovach, 
A.~Honecker, A.~Lascialfari, M.~Leuenberger, and in particular 
B.~Normand for stimulating discussions.

\appendix

\section{Susceptibilities}
\label{sec:a-suscept}

In this appendix, we sketch how the transverse susceptibilities 
$ \chi_{\alpha \alpha}(i \omega_n)$, $ \alpha=y,z$, can be evaluated 
for the FW with Hamiltonian $\hat{H}_0$ [Eq.~(\ref{eq:1-h0})]. With 
$M_y=M_z  = 0$,~\cite{allen:97} 
\begin{eqnarray}
 && \chi_{\alpha \alpha}(\tau) = (g \mu_B)^2 \frac{N \hbar}{4J} 
\delta
(\tau) (1- \langle n_\alpha^2 \rangle) \nonumber \\ &&+ (g \mu_B)^2
\left( \frac{N \hbar}{4J}\right)^2 \frac{1}{Z} \int {\mathcal D}{\bf n}
[i ({\bf n}\times \dot{\bf n})_\alpha + n_\alpha h_x n_x ]_{\tau} \nonumber\\ 
&&
\hspace{1.5cm}
[i ({\bf n}\times \dot{\bf n})_\alpha + n_\alpha h_x n_x ]_0
e^{-\int_0^{\beta \hbar}{\rm d}\tau \, L_0 [{\bf n}]/\hbar},
\label{eq:a1-1}
\end{eqnarray}
where the first (second) square bracket is evaluated at
$\tau$ ($0$). Using the parameterization in Eq.~(\ref{eq:2-param})
and expanding to second order in $\vartheta=\theta-\pi/2$, for $\alpha = y$ 
the square bracket 
reads
\begin{equation}
i ({\bf n}\times \dot{\bf n})_y + n_y h_x n_x = - (h_x - i \dot{\phi}) 
\vartheta \cos \phi - i  \dot{\vartheta} \sin \phi.
\label{eq:a1-2}
\end{equation}
The corresponding expression for $\alpha=z$ can be obtained by setting
$\phi \rightarrow \phi -\pi/2$. Integrating $\vartheta$, we obtain
\begin{eqnarray}
&& \chi_{yy}(\tau) = (g\mu_B)^2 
\frac{N \hbar}{4J}(1-\langle \cos^2 \phi \rangle) \delta(\tau)  \nonumber 
\\ && +
\left( \frac{N \hbar}{4J}\right)^2 \frac{1}{Z}
\int {\mathcal D}\phi \,  e^{-\int_0^{\beta \hbar}{\rm d} \tau \, L_0 [\phi]
/\hbar} 
\nonumber \\ && 
\times [
G_{\vartheta \vartheta}(\tau)  (h_x - i \dot{\phi}(\tau))
\cos \phi (\tau) \,
(h_x - i \dot{\phi})  \cos \phi  \nonumber \\ && 
+
i G_{\vartheta \dot{\vartheta} }(\tau) (h_x - i \dot{\phi}(\tau))
\cos \phi (\tau) \, \sin \phi \nonumber \\ && 
+
i G_{\dot{\vartheta}\vartheta }(\tau) \sin \phi(\tau)
(h_x - i \dot{\phi})  \cos \phi \nonumber \\ && 
+ 
i^2 G_{\dot{\vartheta}\dot{\vartheta}}(\tau) \sin \phi(\tau)
\sin \phi
] .
\label{eq:a1-3}
\end{eqnarray}
The Green functions are defined by $G_{\vartheta \vartheta}(\tau)
= \langle {\rm T}_{\tau} \vartheta (\tau) \vartheta \rangle -
\langle \vartheta \rangle^2$. In the
high-field limit $h_x \gg \omega_0$, all Green functions are strongly peaked 
at 
$\tau = 0$.
Using the adiabatic approximation outlined in Sec.~\ref{sec:0-mqc}, 
we find from $L_0[{\bf n}]$ [Eq.~(\ref{eq:2-action})]
(up to ${\mathcal O}(\omega_0^2/h_x^2)$)
\begin{eqnarray}
G_{\vartheta \vartheta}(\tau) & \simeq & \frac{2 J}{N \hbar (h_x -
i \dot{\phi}) } e^{-(h_x -i \dot{\phi})|\tau|}, \\
G_{\vartheta \dot{\vartheta}}(\tau) &\simeq &
\frac{2 J}{N \hbar} {\rm sgn}(\tau) 
e^{-(h_x -i \dot{\phi})|\tau|}, \\
G_{\dot{\vartheta}\dot{\vartheta}}(\tau) &\simeq& \frac{4J}{N \hbar} \delta(\tau)
- \frac{2 J (h_x - i \dot{\phi}) }{N \hbar } 
e^{-(h_x -i \dot{\phi})|\tau|},
\label{eq:a1-4}
\end{eqnarray}
where $\dot{\phi} = \dot{\phi}(0)$. 
Along the classical path, the field $\phi$ varies on a timescale
$1/\omega_0$, i.e. slowly on the timescale over which the Green 
functions vanish, which allows us to set 
$\exp[-(h_x -i \dot{\phi})|\tau| ]\rightarrow 
(2/(h_x -i \dot{\phi})) \delta(\tau)$ in 
$G_{\vartheta \vartheta}$ and $G_{\dot{\vartheta}\dot{\vartheta}}$.
Because $G_{\vartheta \dot{\vartheta}} = - G_{\dot{\vartheta} \vartheta}$ the
second and third term in Eq.~(\ref{eq:a1-3}) cancel.
To leading order in $\omega_0/h_x$ we then obtain
\begin{eqnarray}
&& \chi_{yy}(\tau) \simeq (g \mu_B)^2 \frac{N \hbar}{4J} \Bigl[
\bigl( 1 - \langle \cos^2 \phi + \sin^2 \phi \rangle \bigr) \delta(\tau)
 \nonumber \\
&& \hspace*{4cm} + \frac{h_x}{2} e^{-h_x |\tau|}  \Bigr]
\nonumber \\
&&= (g \mu_B)^2 \frac{N \hbar h_x}{8J} e^{-h_x |\tau|}  
\simeq (g \mu_B)^2 \frac{N \hbar}{4J} \delta (\tau),
\label{eq:a1-5}
\end{eqnarray}
and
\begin{equation}
\chi_{zz}(\tau) \simeq (g \mu_B)^2 \frac{N \hbar h_x}{8J} e^{-h_x |\tau|}
\simeq (g \mu_B)^2 \frac{N \hbar}{4J} \delta (\tau).
\label{eq:a1-6}
\end{equation}

\section{Two-state model of the modified FW}
\label{sec:a-TSS}

As shown in Secs.~\ref{sec:introduction} and \ref{sec:0-fw}, for weak 
tunneling 
$\tilde{{\mathcal S}}/\hbar \gg 1$, the low-energy sector of the (modified) 
FW can be described 
as a two-state model with basis $|\uparrow\rangle$ and $|\downarrow \rangle$. 
In this approximation, $\hat{H}=-(\tilde{\Delta}/2)[|\uparrow\rangle 
\langle \downarrow| + |\downarrow\rangle \langle \uparrow|]$, where the 
parameters of the original microscopic Hamiltonian enter $\tilde{\Delta}$. 

For any operator $\hat{O}$, the transition matrix element between $|g\rangle$ 
and $|e\rangle$ can be evaluated from
\begin{eqnarray}
\langle e | \hat{O} | g \rangle &=& 
\frac{1}{2} \left( \langle \uparrow | \hat{O} | \uparrow \rangle - 
 \langle \downarrow | \hat{O} | \downarrow \rangle \right. \nonumber \\
&& \hspace*{0.5cm} \left.+ 
\langle \uparrow | \hat{O} | \downarrow \rangle -
 \langle \downarrow | \hat{O} | \uparrow \rangle
\right).
\label{eq:a-transitma}
\end{eqnarray}
For $h_x \gg \tilde{\omega}_0$ the 
state \mbox{$|\uparrow \rangle$}
describes a Gaussian probability distribution for ${\bf n}$ in the plane
$\perp {\bf B}$ with variance $\langle \uparrow | \cos^2 \phi |
\uparrow \rangle = 2 \tilde{J}/N \hbar \tilde{\omega}_0 = 
1/(\tilde{{\mathcal S}}/\hbar )$ [Sec.~\ref{sec:0-mqc}]. Expanding
$\sin \phi \simeq \pm (1 - \cos^2 \phi /2)$ for 
$\phi \simeq \pi/2$ and $\phi \simeq 3 \pi/2$, respectively, we obtain
\begin{eqnarray}
|\langle e | \sin \phi | g \rangle| & \simeq & 
\frac{1}{2} | \langle \uparrow | \sin \phi | \uparrow \rangle - 
 \langle \downarrow | \sin \phi | \downarrow \rangle | \nonumber \\ 
& \simeq & 1 - \frac{\tilde{J}}{N \hbar \tilde{\omega}_0}.
\label{eq:a-transitmb}
\end{eqnarray}
The terms $ \langle \downarrow | \sin \phi | \uparrow \rangle$ are of 
order $\exp[-\tilde{{\mathcal S}}/\hbar]$ and hence negligible in the 
weak tunneling
regime.
Similarly, we also find 
\begin{equation}
|\langle e | \sin^3 \phi | g \rangle |\simeq 
1 - 3\frac{ \tilde{J}}{N \hbar \tilde{\omega}_0}.
\label{eq:a-transitmc}
\end{equation}

\section{Effective Lagrangean for an open AF spin chain}
\label{sec:a-openchain}

In this appendix we show that an open spin chain with an odd number $N-1$
of spins ${\bf s}_i$, $i=2,3,\ldots,N$, can be mapped onto the Lagrangean 
of a modified FW. For simplicity we restrict ourselves to the chain 
\begin{equation}
H = J \sum_{i=2}^{N-1} \hat{{\bf s}}_i \cdot \hat{{\bf s}}_{i+1} 
+ \hbar {\bf h} \cdot \sum_{i=2}^{N} \hat{{\bf s}}_{i}
- k_z \sum_{i=2}^{N} \hat{s}_{i,z}^2 .
\label{eq:openchain-h}
\end{equation}
Again, we assume that the system exhibits AF order, and use the 
staggering ${\bf s}_i=(-1)^{i} s {\bf n} + {\bf l}$ for the
spin fields, with ${\bf l}
\cdot {\bf n} = 0$. Spatial variations of the fields
${\bf n}$ and ${\bf l}$ are strongly suppressed in the small system under
consideration. We hence can proceed as in Sec.~\ref{sec:thermodyn} and carry
out the Gaussian
integral over ${\bf l}$. We obtain
\begin{eqnarray}
&& L_{\rm chain}[{\bf n}]=\frac{(N-1)^2}{8 J (N-2)}
\bigl[ -({\bf h} + h_A n_z {\bf e}_z - i {\bf n}\times \dot{\bf n})^2 
\nonumber 
\\ && + 
(({\bf h} + h_A n_z {\bf e}_z )\cdot {\bf n})^2
\bigr] - (N-1) k_z s^2 n_z^2 \nonumber 
\\ && -s ({\bf h}\cdot {\bf n} - 
i \dot{\phi}(1 + \cos \theta)) ,
\label{eq:openchain-l}
\end{eqnarray}
where $h_A = -2 s k_z/(N-1) \hbar$. Evidently, by appropriate definition of 
$\tilde{J}$
and $\tilde{\omega}_0$, the Lagrangean of the open spin chain can be mapped 
onto that of
the modified FW [Eq.~(\ref{eq:3-lagrange})] with excess spin $\delta s=s$.


\begin{references}
\bibitem{remark1c}Classical spin vectors are denoted without 
hat.
\bibitem{remark1}In order to obtain reliable results for the 
magnetization $M_x(B_x)$ with instanton techniques, one needs 
anisotropies as large as
$k_z = 0.015 J$ ($k_z=0.039 J$) in Fe$_{10}$ (Fe$_{6}$) wheels. These
values are larger than, but of the same order of magnitude as the ones of the
clusters synthesized so far.
\bibitem{remark2}The absolute value of $\delta B_{c,n}$ deviates from 
$4 \tilde{J}/N g \mu_B$ by terms of order $\tilde{\omega}_0^2/h_x^2$, which is
beyond the precision of our result Eq.~(\ref{eq:3-lagrangec}). Hence we 
compare ED and analytical results only for $\delta B_{c,n} (J^\prime) 
- \delta B_{c,n} (J)$, for which this shift nearly cancels.
\bibitem{remark2a}Note that the classical value for $m_z$ at the saddle
points of the path integral, $\phi = \pi/2$ and $3\pi/2$, respectively, 
coincides with the value derived from the classical vector model in 
Sec.~\ref{sec:subl}, as it should.
\bibitem{remark4}Note that also the anisotropy energy favors a spin
configuration as sketched in Fig.~\ref{fig3}c such that the condition 
$h_x \gg |\delta s| 8J/N$ can be relaxed. Consequently, also for Ga dopants a 
tunneling scenario is feasible.
\end{references}
\end{document}